%% file: main.tex
\documentclass[journal]{IEEEtran}
\IEEEoverridecommandlockouts

\usepackage{cite}
\usepackage{amsthm}
\usepackage{amsmath,amssymb,amsfonts,eqnarray}
\usepackage{graphicx}
\usepackage{textcomp}
\usepackage{xcolor}
\usepackage{acronym}
\usepackage{tabularx}
\usepackage{algorithm}
\usepackage{algpseudocode}
\usepackage{multirow}
\usepackage{academicons}
\usepackage{scalerel}
\usepackage{tikz}
\usetikzlibrary{svg.path}
\usepackage{subcaption}
\usepackage{url}
\usepackage{ragged2e}
\usepackage[symbol]{footmisc}
\usepackage{mathtools}
\usepackage{diagbox}
\usepackage{pifont}
\usepackage{booktabs}
\usepackage{siunitx}
\usepackage{tikz,calc}
\usetikzlibrary{tikzmark,decorations.pathreplacing,calc,arrows.meta}
\usepackage{rotating}
\usepackage{dblfloatfix}

\def\BibTeX{{\rm B\kern-.05em{\sc i\kern-.025em b}\kern-.08em
    T\kern-.1667em\lower.7ex\hbox{E}\kern-.125emX}}

  \tikzset{
  orcidlogo/.pic={
    \fill[orcidlogocol] svg{M256,128c0,70.7-57.3,128-128,128C57.3,256,0,198.7,0,128C0,57.3,57.3,0,128,0C198.7,0,256,57.3,256,128z};
    \fill[white] svg{M86.3,186.2H70.9V79.1h15.4v48.4V186.2z}
                 svg{M108.9,79.1h41.6c39.6,0,57,28.3,57,53.6c0,27.5-21.5,53.6-56.8,53.6h-41.8V79.1z M124.3,172.4h24.5c34.9,0,42.9-26.5,42.9-39.7c0-21.5-13.7-39.7-43.7-39.7h-23.7V172.4z}
                 svg{M88.7,56.8c0,5.5-4.5,10.1-10.1,10.1c-5.6,0-10.1-4.6-10.1-10.1c0-5.6,4.5-10.1,10.1-10.1C84.2,46.7,88.7,51.3,88.7,56.8z};
  }
}

\newcommand\orcidicon[1]{\href{https://orcid.org/#1}{\mbox{\scalerel*{
\begin{tikzpicture}[yscale=-1,transform shape]
\pic{orcidlogo};
\end{tikzpicture}
}{|}}}}

\usepackage{hyperref}

\begin{document}
\definecolor{orcidlogocol}{HTML}{A6CE39}

\title{Internet of Drones Simulator: Design, Implementation, and Performance Evaluation}
\author{
\thanks{This work has been submitted to the IEEE for possible publication. Copyright may be transferred without notice, after which this version may no longer be accessible.}
\thanks{This work was partially supported by the Italian MIUR PON projects Pico\&Pro (ARS01\_01061), AGREED (ARS01\_00254), FURTHER (ARS01\_01283), RAFAEL (ARS01\_00305) and by Warsaw University of Technology within IDUB programme (Contract No. 1820/29/Z01/POB2/2021).}
\thanks{G. Grieco, G. Iacovelli, P. Boccadoro and L.A. Grieco are with the Department of Electrical and Information Engineering, Politecnico di Bari, Bari, Italy (email: \textit{name.surname}@poliba.it) and with the Consorzio Nazionale Interuniversitario per le Telecomunicazioni, Parma, Italy.}
	\IEEEauthorblockN{
		Giovanni Grieco~\orcidicon{0000-0002-6326-4244}, ~\IEEEmembership{Graduate Student Member,~IEEE},
		Giovanni Iacovelli~\orcidicon{0000-0002-3551-4584}, ~\IEEEmembership{Graduate Student Member,~IEEE},\\
		Pietro Boccadoro~\orcidicon{0000-0002-7981-9312}, ~\IEEEmembership{Member,~IEEE},
		Luigi Alfredo Grieco~\orcidicon{0000-0002-3443-6924}, ~\IEEEmembership{Senior Member,~IEEE}
	}\\
}

\input{acronyms}

\maketitle

\begin{abstract}
The \ac{IoD} is a networking architecture that stems from the interplay between \acp{UAV} and wireless communication technologies. Networked drones can unleash disruptive scenarios in many application domains. At the same time, to really capitalize their potential, accurate modeling techniques are required to catch the fine details that characterize the features and limitations of \acp{UAV}, wireless communications, and networking protocols.
To this end, the present contribution proposes the \ac{IoD-Sim}, a comprehensive and versatile open source tool that addresses the many facets of the \ac{IoD}.
\ac{IoD-Sim} is an \ac{ns-3}-based simulator organized in a 3-layer stack, composed by (i) the Underlying Platform, which provides the telecommunication primitives for different standardized protocol stacks, (ii) the Core, that implements all the fundamental features of an \ac{IoD} scenario, and (iii) the Simulation Development Platform, mainly composed by a set of tools that speeds up the graphical design for every possible use-case.
In order to prove the huge potential of this proposal, three different scenarios are presented and analyzed from both a software perspective and a telecommunication standpoint.
The peculiarities of this open-source tool are of interest for researchers in academia, as they will be able to extend to model upcoming specifications, including, but not limited to, mobile networks and satellite communications.
Still, it will certainly be of relevance in industry to accelerate the design phase, thus improving the time-to-market of \ac{IoD}-based services.
\end{abstract}


\input{content.tex}

\bibliographystyle{IEEEtran}
\bibliography{bibliography}

\begin{IEEEbiography}
    [{\includegraphics[width=1in,height=1.25in,clip,keepaspectratio]{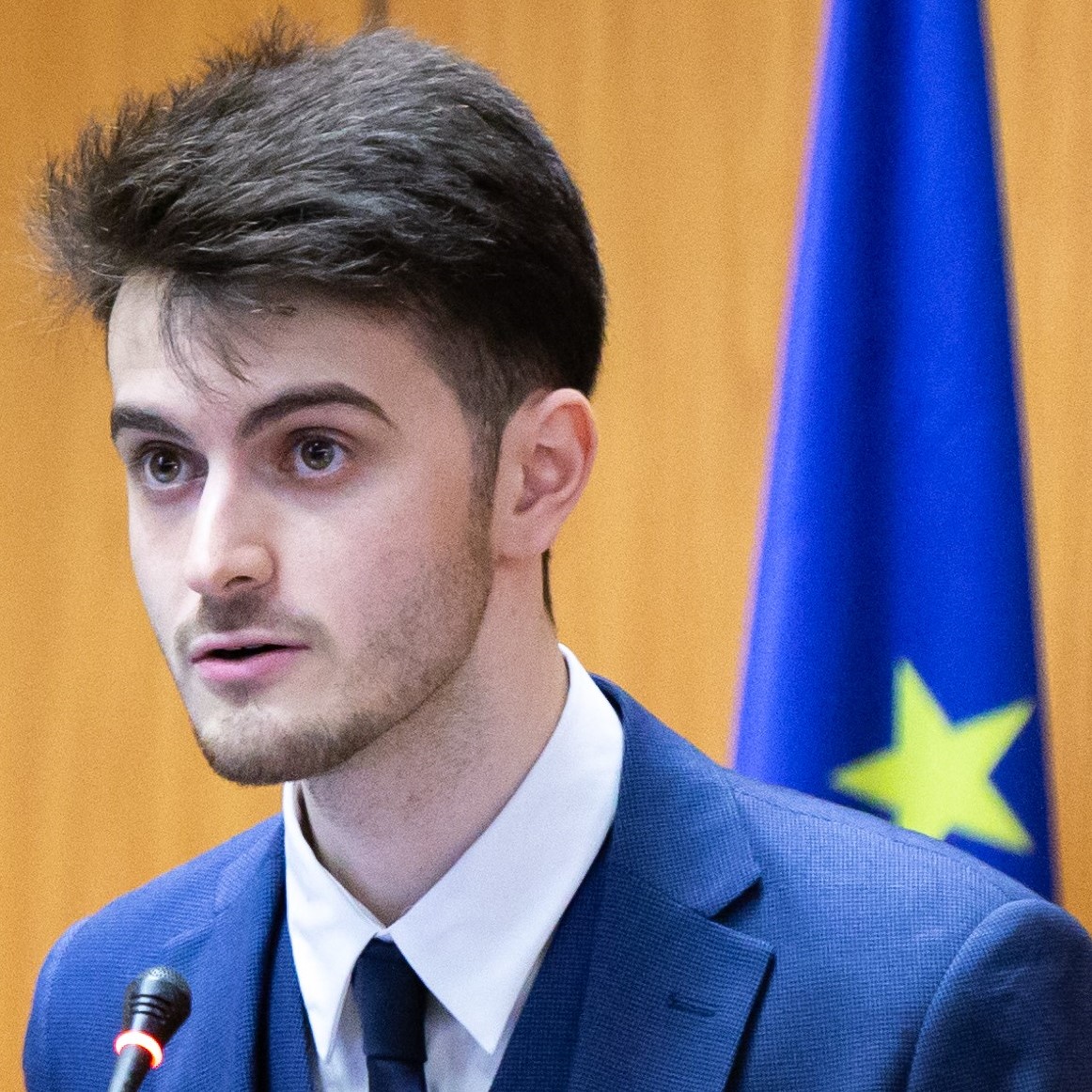}}]{Giovanni Grieco}
received the Dr. Eng. degree (with honors) in Telecommunications Engineering from Politecnico di Bari, Bari, Italy in October 2021. His research interests include Internet of Drones, Cybersecurity, and Future Networking Architectures. He is the principal maintainer of IoD\_Sim. Since 2021 he is a Ph.D. Student at the Department of Electrical and Information Engineering at Politecnico di Bari.
\end{IEEEbiography}
\begin{IEEEbiography}
    [{\includegraphics[width=1in,height=1.25in,clip,keepaspectratio]{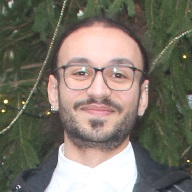}}]{Giovanni Iacovelli}
received the Dr. Eng. degree (with honors) in information engineering from Politecnico di Bari, Bari, Italy, in July 2019. His research interests include Internet of Drones, Machine Learning, Optimization and Telecommunications. Since November 2019 he is a Ph.D. Student at the Department of Electrical and Information Engineering, Politecnico di Bari.
\end{IEEEbiography}
\begin{IEEEbiography}
    [{\includegraphics[width=1in,height=1.25in,clip,keepaspectratio]{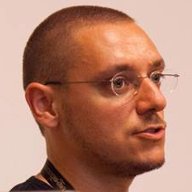}}]{Pietro Boccadoro}
received the Dr. Eng. degree (with honors) in electronic engineering from Politecnico di Bari, Bari, Italy, in July 2015. From Nov. 2015 to Oct. 2020, he collaborated as a researcher at Politecnico di Bari. In 2021, he finished his Ph.D. Course. He is cureently R\&D Software engineer at Nextome srl. His research interests include Internet of Drones, Robotic-aided IoT and Future Internet Architectures.
\end{IEEEbiography}
\begin{IEEEbiography}
    [{\includegraphics[width=1in,height=1.25in,clip,keepaspectratio]{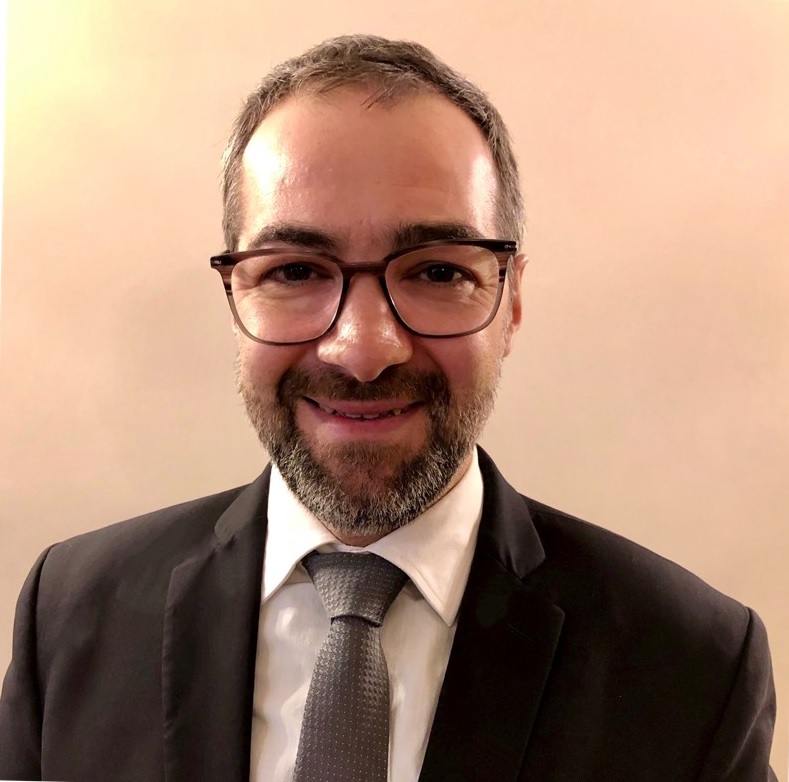}}]{L. Alfredo Grieco}
is a full professor in telecommunications at Politecnico di Bari. His research interests include Internet of Things, Future Internet Architectures, and Nano-communications. He serves as Founder Editor in Chief of the Internet Technology Letters journal (Wiley) and as Associate Editor of the IEEE Transactions on Vehicular Technology journal (for which he has been awarded as top editor in 2012, 2017, and 2020).
\end{IEEEbiography}

\end{document}

%% file: acronyms.tex
\acrodef{5G}{Fifth Generation}
\acrodef{6lo}{\ac{IPv6} over networks of resource-constrained nodes}
\acrodef{6LoWPAN}{\ac{IPv6} over Low power Wireless Personal Area Networks}
\acrodef{6TiSCH}{IPv6 over the TSCH mode of IEEE 802.15.4}
\acrodef{6top}{6tisch Operation Sublayer}

\acrodef{A2A}{Air-to-Air}
\acrodef{A2G}{Air-to-Ground}
\acrodef{ABP}{Activation By Personalization}
\acrodef{ACK}{Acknowledgment}
\acrodef{AES}{Advanced Encryption Standard}
\acrodef{AGV}{Automated Guided Vehicle}
\acrodef{AMQP}{Advanced Message Queuing Protocol}
\acrodef{AP}{Allocation Policy} 
\acrodef{API}{Application Programming Interface}
\acrodef{AQ}{Autonomous Quadcopter}
\acrodef{ASAP}{As-Soon-As-Possible Schedule Another Parent}
\acrodef{ASN}{Absolute Slot Number}
\acrodef{AST}{Abstract Syntax Tree}
\acrodef{ASV}{Autonomous Surface Vehicle}
\acrodef{ATC}{Air Traffic Control}
\acrodef{AUV}{Autonomous Underwater Vehicle}

\acrodef{BS}{Base Station}
\acrodef{BSS}{Basic Service Set}
\acrodef{BVLoS}{Beyond Visual Line of Sight}
\acrodef{BW}{Bandwidth}

\acrodef{CH}{Cluster Head}
\acrodef{CCTV}{Closed Circuit TeleVision}
\acrodef{CEA}{Cell Estimation Algorithm}
\acrodef{CIR}{Channel Impulse Response}
\acrodef{CoAP}{Constrained Application Protocol}
\acrodef{COM}{Common Object Model}
\acrodef{CPS}{Cyber-Physical System}
\acrodef{CPU}{Central Processing Unit}
\acrodef{CR}{Cognitive Radio}
\acrodef{CRC}{Cyclic Redundancy Check}
\acrodef{CSMA}{Carrier Sense Multiple Access}
\acrodef{CSMA/CA}{Carrier Sense Multiple Access with Collision Avoidance}
\acrodef{CSN}{Community Seismic Network}
\acrodef{CSS}{Chirp Spread Spectrum}

\acrodef{D2D}{Device-to-Device}
\acrodef{DAG}{Directed Acyclic Graph}
\acrodef{DAO}{Destination Advertisement Object}
\acrodef{DCL}{Drone Control Layer}
\acrodef{DETB}{Dynamic Energy \& Traffic Balance}
\acrodef{DGAC}{Direction Générale de l'Aviation Civile}
\acrodef{DHCP}{Dynamic Host Configuration Protocol}
\acrodef{DIO}{Destination Information Object}
\acrodef{DNS}{Domain Name System}
\acrodef{DOC}{Drone Orchestration Center}
\acrodef{DODAG}{Destination Oriented DAG}
\acrodef{DSSS}{Direct Sequence Spread Spectrum}
\acrodef{DRY}{Don't Repeat Yourself}

\acrodef{E2E}{End-to-End}
\acrodef{EASA}{European Aviation Safety Agency}
\acrodef{EbI}{Eastbound Interface}
\acrodef{ENAC}{Ente Nazionale per l'Aviazione Civile}
\acrodef{EOF}{End Of Frame}
\acrodef{ePFC}{extended Potential Field Controller}

\acrodef{FAA}{Federal Aviation Administration}
\acrodef{FANET}{Flying Ad-hoc NETwork}
\acrodef{FBS}{Flying Base Station}
\acrodef{FCC}{Flight Control Center}
\acrodef{FEC}{Forward Error Correction}
\acrodef{FSM}{Finite State Machine}
\acrodef{FSO}{Free Space Optical}
\acrodef{FPU}{Floating Point Unit}

\acrodef{GBS}{Ground Base Station}
\acrodef{GCS}{Ground Control Station}
\acrodef{GMP}{Gateway Message Protocol}
\acrodef{GDI}{Global Drone Identifier}
\acrodef{GIS}{Geographic Information System}
\acrodef{GPS}{Global Position System}
\acrodef{GRI}{Global Role IDentifier}
\acrodef{GU}{Ground User}
\acrodef{GUI}{Graphical User Interface}

\acrodef{HAL}{High-level Abstraction Layer}
\acrodef{HitL}{Hardware in the Loop}
\acrodef{HTTP}{HyperText Transfer Protocol}
\acrodef{KPI}{Key Performance Index}
\acrodef{KPIs}{Key Performance Indices}

\acrodef{IANA}{Internet Assigned Numbers Authority}
\acrodef{ICN}{Information-Centric Networking}
\acrodef{ICT}{Information and Communication Technologies}
\acrodef{IETF}{Internet Engineering Task Force}
\acrodef{IIoT}{Industrial Internet of Things}
\acrodef{IMC}{Internal Model Control}
\acrodef{IMU}{Inertial Measurement Unit}
\acrodef{IoD}{Internet of Drones}
\acrodef{IoT}{Internet of Things}
\acrodef{IP}{Internet Protocol}
\acrodef{IPC}{Inter-Process Communication}
\acrodef{IPv6}{Internet Protocol version 6}
\acrodef{IR}{Intermediate Representation}
\acrodef{ISM}{Industrial, Scientific and Medical}
\acrodef{ITS}{Intelligent Transportation System}
\acrodef{IDE}{Integrated Development Environment}
\acrodef{IPv4}{Internet Protocol version 4}
\acrodef{IMSI}{International Mobile Subscriber Identity}

\acrodef{JSON}{JavaScript Object Notation}

\acrodef{LoRa}{Long Range}
\acrodef{LoS}{Line of Sight}
\acrodef{LPWAN}{Low Power Wide Area Network}
\acrodef{SID}{Swarm Identifier}
\acrodef{LTE}{Long-Term Evolution}
\acrodef{LxC}{Linux Containers}

\acrodef{M2M}{Machine-to-Machine}
\acrodef{MAC}{Medium Access Control}
\acrodef{MANET}{Mobile Ad-hoc NETwork}
\acrodef{MAVLink}{Micro Air Vehicle Link}
\acrodef{MCU}{Micro-Controller Unit}
\acrodef{MSPRT}{Multi Sequential Probability Ratio Test}
\acrodef{MTC}{Machine-Type Communication}
\acrodef{mcMTC}{mission critical Machine-Type Communication}
\acrodef{MTU}{Maximum Transmission Unit}
\acrodef{MEMS}{Micro Electro-Mechanical Systems}
\acrodef{MIC}{Message Integrity Code}
\acrodef{MMI}{Modified Mercalli Intensity}
\acrodef{MPwise}{Multi-Parameter Wireless Sensing System}
\acrodef{MQTT}{Message Queue Telemetry Transport}

\acrodef{NAT}{Network Address Translation}
\acrodef{NB-IoT}{Narrow Band IoT}
\acrodef{NbI}{Northbound Interface}
\acrodef{NIC}{Network Interface Controller}
\acrodef{NLoS}{Non-Line of Sight}
\acrodef{ns-3}{Network Simulator 3}

\acrodef{O-QPSK}{Offset-Quadrature Phase-Shift Keying}
\acrodef{OS}{Operating System}
\acrodef{OSI}{Open Systems Interconnection}
\acrodef{OTF}{On The Fly}

\acrodef{PaaS}{Platform as a Service}
\acrodef{PCH}{Precompiled Header}
\acrodef{PDU}{Protocol Data Unit}
\acrodef{PGA}{Peak Ground Acceleration}
\acrodef{PGD}{Peak Ground Displacement}
\acrodef{PGHA}{Peak Ground Horizontal Acceleration}
\acrodef{PGV}{Peak Ground Velocity}
\acrodef{PGVA}{Peak Ground Vertical Acceleration}
\acrodef{PoI}{Point of Interest}
\acrodef{PpoI}{Points of Interest}
\acrodef{pub/sub}{publish/subscribe}
\acrodef{P2P}{Peer-to-Peer}
\acrodef{PHY}{Physical Layer}
\acrodef{PID}{Proportional-Integral-Derivative}

\acrodef{QoE}{Quality of Experience}
\acrodef{QoL}{Quality of Link}
\acrodef{QoS}{Quality of Service}

\acrodef{RAN}{Radio Access Network}
\acrodef{RDBMS}{Relational DataBase Management System}
\acrodef{RoI}{Region of Interest}
\acrodefplural{RoI}[RoIs]{Regions of Interest}
\acrodef{ROLL}{Routing Over Low power and Lossy networks}
\acrodef{ROS}{Robot Operating System}
\acrodef{ROV}{Remotely Operated underwater Vehicle}
\acrodef{RPL}{Routing Protocol for Low-power and Lossy networks}
\acrodef{RSSI}{Received Signal Strength Indicator}
\acrodef{RTT}{Round Trip Time}

\acrodef{SbI}{Southbound Interface}
\acrodef{SF}{Spreading Factor}
\acrodef{SDN}{Software Defined Network}
\acrodef{SIM}{Subscriber Identity Module}
\acrodef{SLAM}{Simultaneous Localization And Mapping}
\acrodef{SINR}{Signal-to-Interference-plus-Noise Ratio}
\acrodef{SPI}{Serial Peripheral Interface}
\acrodef{SSID}{Service Set IDentifier}

\acrodef{TCP}{Transmission Control Protocol}

\acrodef{UDP}{User Datagram Protocol}
\acrodef{UAV}{Unmanned Aerial Vehicle}
\acrodef{UML}{Unified Modeling Language}
\acrodef{UTM}{Unmanned Aerial System Traffic Management}
\acrodef{UI}{User Interface}

\acrodef{VANET}{Vehicular Ad-hoc NETwork}
\acrodef{VLC}{Visible Light Communication}
\acrodef{VLoS}{Visual Line Of Sight}
\acrodef{VM}{Virtual Machine}
\acrodef{VPL}{Visual Programming Language}
\acrodef{VPN}{Virtual Private Network}

\acrodef{WbI}{Westbound Interface}

\acrodef{XML}{Extensible Markup Language}
\acrodef{XSF}{X4 Stationnary Flyer}
\acrodef{LSA}{Logical Swarm Address}

\acrodef{ZSP}{Zone Service Provider}
\acrodef{IoD-Sim}{Internet of Drones Simulator}
\acrodef{GSL}{GNU Scientific Library}
\acrodef{PLR}{Packet Loss Ratio}

%% file: content.tex
\section{Introduction}\label{sec:introduction}
The \acf{IoD} \cite{GBW16} is one of the hottest research topic in telecommunications today \cite{BOCCADORO2021102600}. At first, it might appear as an extension of the \ac{IoT}, with \acfp{UAV} playing the role of smart objects able to fly. Nevertheless, in the \ac{IoD}, drones are tasked to accomplish mission plans with multiple objectives. Since they can also fly in organized groups, namely swarms, it is worth remarking that they are able to continuously optimize their trajectory, and coordinate among themselves.  Drones are currently involved in the delivery of value-added services in many applications, including goods delivery, environmental surveying, first-aid units in disruptive events \cite{BOCCADORO2021102600,LBD+21}, and \ac{FBS} in \ac{5G} \& Beyond scenarios, with multiple users requesting connectivity at the same time and in the same area \cite{BOCCADORO2021102600,AMA+19,LBD+21}.
All this turned the \ac{IoD} from a niche subject to a mainstream research topic in networking.
It must be noted that the adoption of drones in industry is also a huge commercial opportunity, as testified by the several billions forecasts already available for multiple business sectors \cite{BOCCADORO2021102600}.

Even though several applications are now including drones, and they may look like off-the-shelf utilities, the design of complex \ac{IoD} systems still requires advanced methodologies to effectively unleash the potential of services based on networked drones. In 5G \& Beyond scenarios, ubiquitous connectivity and relaying capabilities are required to interact with both terrestrial entities, i.e., ground \acp{BS} and users, as well as aerospace ones, such as satellites.
In this regard, channel capacity, available/required data rates, dedicated bandwidth and frequencies, must be characterized, bearing in mind that every link may be realized with a different telecommunication protocol.
Moreover, given the variety of available drones on the market, an accurate suitability assessment based on their specifics is required.

\begin{table*}[!htbp]
    \caption{Summary of the comparison of the available solutions}
    \label{tab:comparison}
    \centering
    \begin{tabular}{l|c|c|c|c|c|c|c|c|}
    \cline{2-9}  & \textbf{\cite{BSL18}} & \textbf{\cite{ZTS+17}} & \textbf{\cite{marconato2017avens}} & \textbf{\cite{MAC+19}} & \textbf{\cite{TFD+20}} & \textbf{\cite{acharya2021cornet}} & \textbf{\cite{PLL+20}} & \textbf{\ac{IoD-Sim}} \\ \hline
    \multicolumn{1}{|l|}{\textbf{Open source code}} &  &  &  & \checkmark &  & \checkmark &  & \checkmark \\ \hline
    \multicolumn{1}{|l|}{\textbf{Modularity}} & \checkmark & \checkmark & \checkmark &  &  &  &  & \checkmark \\ \hline
    \multicolumn{1}{|l|}{\textbf{Scalability}} &  & \checkmark & \checkmark &  &  &  &  & \checkmark \\ \hline
    \multicolumn{1}{|l|}{\textbf{Visual scenario configuration}} & \checkmark & \checkmark &  &  &  & \checkmark &  &  \checkmark \\ \hline
    \multicolumn{1}{|l|}{\textbf{Network simulations}} &  & \checkmark & \checkmark &  & \checkmark & \checkmark & \checkmark & \checkmark \\ \hline
    \multicolumn{1}{|l|}{\textbf{Support multiple networking standards}} &  & \checkmark &  &  & \checkmark &  &  & \checkmark \\ \hline
    \multicolumn{1}{|l|}{\textbf{Multi-stack protocols support}} & \checkmark &  &  &  &  &  &  & \checkmark \\ \hline
    \multicolumn{1}{|l|}{\textbf{Graphically-assisted trajectory design}} & \checkmark & \checkmark &  &  &  &  &  & \checkmark \\ \hline
    \multicolumn{1}{|l|}{\textbf{Aerodynamics simulations}} & \checkmark & \checkmark &  & \checkmark &  &  &  &  \\ \hline
    \multicolumn{1}{|l|}{\textbf{Power consumption models}} &  &  &  &  & \checkmark &  &  & \checkmark \\ \hline
    \multicolumn{1}{|l|}{\textbf{\ac{HitL} support}} & \checkmark & \checkmark &  &  &  &  &  & \\ \hline
    \multicolumn{1}{|l|}{\textbf{High-level application development support}} &  & \checkmark &  & \checkmark & \checkmark &  & \checkmark & \checkmark \\ \hline
    \multicolumn{1}{|l|}{\textbf{Ready-to-use \ac{IoD} applications}} &  &  &  &  &  &  &  & \checkmark \\ \hline
    \multicolumn{1}{|l|}{\textbf{Machine-readable results}} &  & \checkmark & \checkmark & \checkmark &  &  &  & \checkmark \\ \hline
    \multicolumn{1}{|l|}{\textbf{Human-readable results}} & \checkmark & \checkmark & \checkmark & \checkmark &  &  & \checkmark &  \checkmark \\ \hline
\end{tabular}
\end{table*}

Differently from the available \ac{IoD} simulators \cite{BSL18, ZTS+17, marconato2017avens, MAC+19, TFD+20, acharya2021cornet, PLL+20}, which do not cover all the aforementioned aspects, this work proposes a comprehensive open-source simulation platform, namely \ac{IoD-Sim}\footnote{\url{https://github.com/telematics-lab/IoD_Sim}} \cite{IODSIM}.
Since it was first released, it has been sensibly modified, partially re-written, and thoroughly refactored in order to create complex operating scenarios.
The architecture is designed as a 3-layer stack: (i) the \textit{Underlying Platform}, which includes a set of technologies and libraries able to perform high-precision numerical computation; (ii) the \textit{Core}, which embeds a set of unique \ac{IoD}-related features; (iii) the \textit{Simulation Development Platform} that allows high-level mission design and analysis of simulation results.
\ac{IoD-Sim} is able to create realistic simulations by extending the available features of \ac{ns-3} to address the relevant aspects of the \ac{IoD}, thus including mission design, trajectory planning, hardware and application configuration, mobile wireless communications, mobility and energy consumption models, on-board peripherals, and integration with other network entities.

To prove its potential and to validate its manifold functionalities, an extensive and diversified simulation campaign is carried out. Three different scenarios are conceived through the proposed high-level mission design tool, which grants a welcoming user experience via a convenient interface.
The different scenarios are characterized in terms of network topologies, communication technologies, drones' equipment, and software applications. In particular, signals experience different propagation conditions introduced by the adoption of channel models that vary from ideal conditions, i.e., free space, up to more realistic ones, i.e., densely populated urban environments.
Lastly, simulation results are analyzed to obtain relevant \acp{KPI}, such as \ac{SINR}, throughput, power consumption, latency, and \ac{PLR}, from which thoughtful insights are derived.

The present contribution is structured as follows:
Section \ref{sec:related-works} summarizes the reference state of the art.
Section \ref{sec:architectural-overview} presents a general overview of the architecture of the simulator.
Section \ref{sec:underlying_platform} describes the underlying platform and the rational for its choice.
Section \ref{sec:sim_core} discusses the core of the simulator in detail, with dedicated subsections about the main building blocks of the project. A thorough explanation of the involved mobility models is given together with all the supported communication technologies, and the involved logical entities.
Section \ref{sec:interfaces} focuses on the simulation design, discussing the detail of helpers, thus explaining the role and importance of scenario configurations.
Section \ref{sec:simulations} is dedicated to the simulation campaign; after an initial focus on scenarios description, the outcomes are discussed to highlight the main findings.
Finally, Section \ref{sec:conclusions} concludes the work and draws future work possibilities.

\section{Related Works}\label{sec:related-works}
To improve, and speed up, both the design and the prototyping phases of \ac{IoD} systems, simulations are widely conceived as a useful aid. Even though simulating drones is a challenging task, it has been dealt with by many contributions so far \cite{BSL18,ZTS+17,marconato2017avens,MAC+19,TFD+20,acharya2021cornet,PLL+20}.
Overall, these works approach \ac{IoD} simulations from two different points of view. The first focuses on the dynamics of the flight, thus including mechanical energy and kinetics.
These works employ \ac{ROS} \cite{quigley2009ros} and Gazebo \cite{1389727} as base platforms \cite{MAC+19,acharya2021cornet}.
The second, instead, focuses on accurate drones networking simulations \cite{BSL18,ZTS+17,marconato2017avens,TFD+20,PLL+20}, mainly based on \ac{ns-3} \cite{Riley2010} and OMNeT++ \cite{Varga2010}, in which \acp{UAV} are envisioned as nodes exchanging data with certain frequencies using well-known protocols belonging to wireless networks, that can either be cellular or Wi-Fi.

The contribution presented in \cite{BSL18} models \acp{UAV} and discusses their functionalities and possible applications. In particular, the proposal introduces FlyNetSim, a software that aims at simulating not only flight operations but also networking communication primitives and principles. The simulator can work with group of drones operating together in a reference ecosystem. The most interesting functionalities are: (i) UAV control over Wi-Fi, (ii) multi-network communications, (iii) \ac{D2D} communications for swarms, and (iv) \ac{IoT} and data streaming.

In \cite{ZTS+17}, instead, it is proposed CUSCUS, a simulation architecture for control systems in the context of drones' networks. The proposal is able to simulate the mechanisms for the control of drones operations and it is claimed to be highly flexible and scalable. The proposed simulator leverages the \ac{ns-3} capabilities to work with virtual interfaces simulating real time systems, eventually composed by swarms.

\cite{marconato2017avens} presents AVENS, which is a hybrid network simulation framework specifically designed to evaluate the performance of intelligent aerial vehicles. Here, drones are communicating using some of the most well-known communication protocols for \acp{FANET}. Differently from other contributions, AVENS is focused on modelling realistic flight condition aspects. On top of that, it uses a layered architecture that acts as an interpreter and code generator, namely LARISSA, thanks to specified simulation parameters and settings. All the results are obtained by the integration and interoperability with the OMNeT++ simulator.

The proposal in \cite{MAC+19} is an interesting simulation framework for unmanned aircraft systems traffic management. It leverages both \ac{ROS} and Gazebo to implement high-level flight services. The simulator is an interesting solution for prototyping missions and controlling both rotary and fixed-wing drones flying in the same environment.

The work presented in \cite{TFD+20} discusses a Java-based simulation framework for \ac{FANET} networks and their applications. In particular, it models the coverage area of each device in the scenario. At the same time, it considers a mobility model for ground entities, i.e., humans in the operating area. Drones’ characterization is herein discussed in terms of limited autonomy and battery recharging needs. To reach this aim, an energy consumption model has been included to evaluate the footprint associated with the flight of a drone. For the sake of completeness, it must be said that this work neglects the contributions due to collision issues and consequent behavior.

\cite{acharya2021cornet} proposes CORNET 2.0, a middleware to simulate Robots, in general, both in physical and networking contexts. It reaches the aim of designing a path planning solution that is simulated by Gazebo and Mininet-WiFi.

The work presented in \cite{PLL+20} proposes a discrete-time, event-based, co-simulation scheme for networks composed by multiple drones, also configured in swarms. The simulator can carry out both flight and network simulations. This solution is of interest because there is intrinsic codependency between the flight status and the networking operations carried out by each drone in the scenario. This contribution is of relevance because it is claimed to guarantee reliability and real-time availability thanks to the possible integration of existing simulators.
This work claims that other available simulators do not implement realistic and reliable mobility models for drones.

When considering all the aforementioned contributions, only the latter shows some similarities with \ac{IoD-Sim} \cite{IODSIM}. For example, both of them operate as discrete-time and event-driven simulators. Nevertheless, it is worth noting that the discrete-time operating mode of \cite{IODSIM} is motivated by the adoption of \ac{ns-3} \cite{Riley2010} as a core network simulator. Another aspect is related with the synthetic trajectories that are implemented in \ac{IoD-Sim}, that are described by closed-form mathematical expressions. Hence, in case the network simulator could be substituted, the mobility models provided by \ac{IoD-Sim} could be used even with continuous-time simulators.

In Table \ref{tab:comparison}, a comparison is made about the main characteristics and features of the referenced works.

\section{Architectural Overview}\label{sec:architectural-overview}
\begin{figure}[htbp]
    \centering
    \includegraphics[width=\columnwidth]{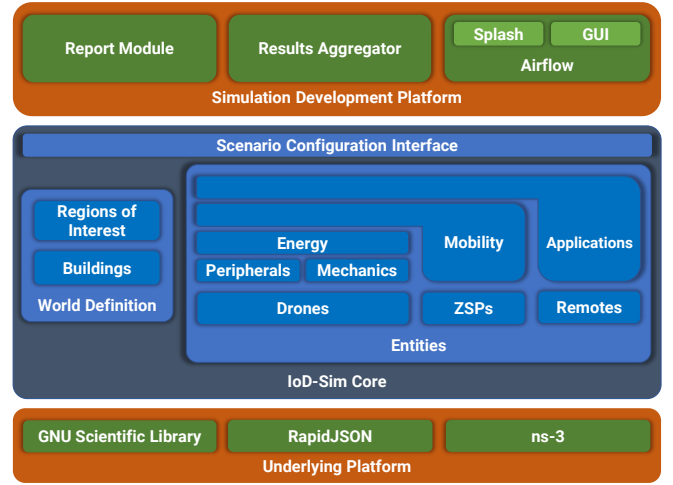}
    \caption{Overview of \ac{IoD-Sim} Architecture.}
    \label{fig:iodsim-architecture}
\end{figure}

The architecture of \ac{IoD-Sim} (see Figure \ref{fig:iodsim-architecture})
is organized into three parts: (i) the \textit{Underlying Platform}, that provides the necessary networking components and performs advanced mathematical and parsing operations, (ii) the \textit{Core} of the simulator, which implements the foundation of \ac{IoD}-related features, and (iii) the so called \textit{Simulation Development Platform}, a high-level component which allows to develop, configure, and analyze advanced scenarios.
Each part is modular by design and proposes peculiar functionalities that are depicted as blocks and described in what follows.
The joint adoption of these components enables different simulation scenarios, starting from a highly flexible and general purpose one, which is configurable by higher-level entities.

In particular, the \ac{IoD-Sim} core, as will be discussed in Section \ref{sec:sim_core}, details all the aspects related to \ac{IoD} entities, especially drones and \acp{ZSP}, spanning from their mechanics and peripherals to remote information services.
Moreover, the whole simulated environment is described by a set of elements, such as specific areas of interest, as well as the presence of buildings in the reference scenario.

With reference to the Simulation Development Platform part of the architecture, it is worth specifying that \ac{IoD-Sim} includes Airflow, a high-level visual configuration environment that drastically eases the interaction between the user and the simulator, i.e., scenario set-up and management.

Finally, a report module, that will be analyzed in Section \ref{sec:report_module}, guarantees the readability of simulation results in a clear XML schema, which eases data processing.
This eases the integration of the \ac{IoD-Sim} with external, e.g., third-party, tools specifically designed to create scenario configurations by customising both parameters (e.g., mechanical properties of the drones) and characteristics (e.g., employed communication technologies).

\section{Underlying Platform}\label{sec:underlying_platform}
The \textit{Underlying Platform} is a foundation composed by the \ac{GSL}, \textit{RapidJSON}, and \ac{ns-3}.

\ac{GSL} is a numerical computing framework which implements numerous routines and low-level data structures, such as complex numbers, linear algebra, data analysis, and interpolation. Furthermore, it is offered in Linux-derived distributions with first-class support \cite{galassi2002gnu}.

\textit{RapidJSON} is a parser and generator of \ac{JSON} code. It is one of the most adopted \ac{JSON} libraries available for C++ projects. It eases the creation, traversal, validity check, and analysis of \ac{JSON} codes \cite{yip2015rapidjson}.
\textit{RapidJSON} has been chosen for its high performance and its extensive and flexible high-level \acp{API}.

Finally, \ac{ns-3} emerges as the most relevant component: it is a solid and mature discrete-time event-based network simulator. \ac{ns-3} is an open-source project that provides a solid simulation engine and various models for network design and testing. Started in 2006, it is a collection of different C++ and Python objects that implements several aspects of networking elements. The fundamental building block of \ac{ns-3} is \texttt{ns3::Node}, an abstract object which represents a generic host in a network. It can be aggregated with other objects and models, e.g., the common TCP/IP stack over Ethernet, to simulate networking behaviour.
Other interesting features in \ac{ns-3} are (i) \texttt{ns3::Channel}, which simulates the communication channel between \texttt{ns3::Node} objects, (ii) \texttt{ns3::NetDevice}, which represents the node networking interface, and (iii) \texttt{ns3::Application}, which sits on top of the protocol stack to produce or consume high-level information.

Furthermore, a \texttt{ns3::Node} can be aggregated with \textit{Mobility Models}, \textit{Energy Consumption Models}. This possibility is not limited to those models, since the support can be extended to any other model that adds new features beyond basic networking. To this end, nodes have the potential to move in space and, hence, drain current from a \texttt{ns3::EnergySource}.

Besides, traces and probes allow to track and record simulation data in log files that are typically encoded in textual ASCII, or PCAP.
In a nutshell, \ac{IoD-Sim} treats \ac{ns-3} as a foundation, extending it with new features that are focused on accurate drone simulations, mobile wireless communications, energy consumption, and their integration with on-board peripherals and ground communication infrastructures.

\section{Core of IoD-Sim}\label{sec:sim_core}
This Section presents the building blocks of the \ac{IoD-Sim} Core, which is the main part of the simulator.

A simulation scenario requires the definition of a simulated world, described by \acp{RoI} and buildings. In this world, entities, i.e., drones, \acp{ZSP}, and remotes, are simulated in a network topology defined by a set of communication models. Each drone is characterized by a mission plan defined by a set of points of interests, which in turn describe a curvilinear trajectory. Furthermore, a drone can be equipped with an energy consumption model, which relies on a set of mechanical properties, and a set of peripherals. Entities in general can host one or more communication stacks and applications. While drones and \acp{ZSP} are connected together according to the configuration of the \ac{IoD} infrastructure, remotes are reachable through a backbone that simulates the Internet behavior. All these blocks are configurable through an abstraction interface focused on interpreting a high-level description of the scenario encoded in \ac{JSON} format.

\subsection{World Definition}
\ac{IoD-Sim} offers the possibility to define parameters related to the simulated world, i.e., the environment in which the simulation takes place. The two main features are the buildings and the Regions of Interest.

The virtual world in \ac{IoD-Sim} is a theoretically infinite space. The space can be filled with entities, which could be Drones, \acp{ZSP} and \textit{Remotes}, but also with \acp{RoI} and \textit{Buildings}.

\subsubsection{Buildings}
\begin{figure}
    \centering
    \includegraphics[width=\columnwidth]{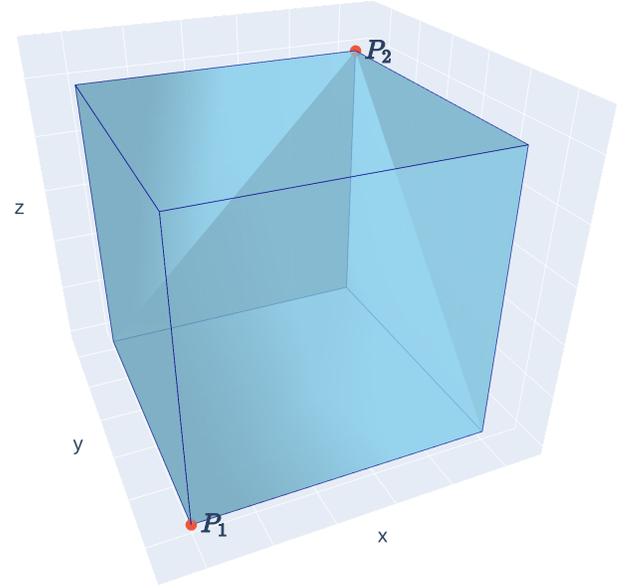}
    \caption{Example of box placement with two points, $P_1$ and $P_2$, in order to create a Building or a \acs{RoI} in the simulated world.}
    \label{fig:exmaple-box-placement}
\end{figure}

The virtual world can be enriched with obstacles, i.e., \textit{Buildings}. They are used to represent urban scenarios, thus making simulations that are particularly suitable for research in \textit{Smart Cities}. \ac{IoD-Sim} provides an abstraction layer to configure and place buildings in the virtual world, relying on \texttt{ns3::BuildingsHelper} and \texttt{ns3::Building} objects. A \texttt{ns3::Building} is a collisionless 3D object with the following properties:
\begin{itemize}
    \item \texttt{boundaries}, which defines the box dimension in the space. Boundaries can be defined by an array of two points organized as $[P^{(1)}_x,\,P^{(2)}_x,\,P^{(1)}_y,\,$ $P^{(2)}_y,\,P^{(1)}_z,\,P^{(2)}_z]$. A representation of these two points is given in Figure \ref{fig:exmaple-box-placement}.
    \item \texttt{type} of building, which can be either \texttt{commercial}, \texttt{residential}, or \texttt{office}.
    \item type of \texttt{walls} material, which can be \texttt{wood}, \texttt{concreteWithWindows}, \texttt{concreteWithoutWindows}, and \texttt{stoneBlocks}.
    \item number of \texttt{floors}.
    \item number of \texttt{rooms} along the x and y axis, per floor. The rooms are placed in a grid position.
\end{itemize}
Such a feature is important for what concern \ac{LTE} communication fading, which varies according to the characteristics of each building.

\subsubsection{\aclp{RoI}}
A \texttt{ns3::InterestRegion} is a 3D box placed on the simulated world defined, as for buildings, by a vector of two points.
Throughout the simulation, it is possible to retrieve and to update the current set of coordinates with \texttt{GetCoordinates()} and \texttt{SetCoordinates()} methods, respectively.

The whole set of these areas is managed by \texttt{ns3::InterestRegionContainer}, which helps to create \acp{RoI} and group them. This utility object provides a (i) \texttt{Create()} method to generate and index \acp{RoI}, (ii) \texttt{GetN()} to report the number of created regions, (iii) \texttt{GetRoi()} to retrieve the $i^\text{th}$, and (iv) \texttt{Begin()} and \texttt{End()} iterators to traverse the entire container.
Moreover, the \texttt{InterestRegionContainer::IsInRegions()} method acknowledges the presence of a drone in multiple areas, thus granting the possibility to trigger specific events during the simulation. For instance, \textit{Drone} operations can be restricted to a limited space, leading to an optimization of \textit{Drone} power consumption.

\subsection{Drones}
\begin{table}
    \caption{\texttt{ns3::Drone} properties in \ac{IoD-Sim}.}
    \label{tab:drone_properties}
    \centering
    \begin{tabular}{ll}
        \toprule
        Name & Unit of Measurement \\
        \midrule
        Mass & $\si{kg}$ \\
        Rotor Disk Area & $\si{m^2}$ \\
        Drag Coefficient & \textit{(dimensionless)} \\
        Peripherals & \\
        \midrule
        Weight Force & $\si{N}$ \\
        Air Density & $\si{kg/m^3}$ \\
        \bottomrule
    \end{tabular}
\end{table}

\ac{IoD-Sim} provides \texttt{ns3::Node} derivatives to consider the characteristics of key actors commonly found in an \ac{IoD} simulation. The \texttt{ns3::Drone} class characterizes a rotary-wing \ac{UAV} and it is registered as a new \texttt{TypeId} in \ac{ns-3}, along with its mechanical properties, shown in Table \ref{tab:drone_properties}.
While the first four properties can be defined by the user, the last two are a direct consequence of the given characterization.
\texttt{ns3::Drone} properties can be set by means of \ac{ns-3} attributes or by its public object interface. Its mass can be updated any time by means of \texttt{SetMass()}. Upon update, the drone weight force is also updated in cascade by multiplying the new mass with the constant gravity acceleration.
The rotor disk area and its drag coefficient can be updated in the same manner by means of \texttt{SetArea()} and \texttt{SetDragCoefficient()} methods, respectively.
Furthermore, \texttt{ns3::Drone} properties can always be read any time during the simulation through \ac{ns-3} attributes and object getters, such as \texttt{GetMass()}, \texttt{GetWeight()}, \texttt{GetArea()}, and \texttt{GetDragCoefficient()}.

Drones can be grouped together in \texttt{ns3::DroneContainer} and can be statically referenced by their unique identifier in the simulation through \texttt{ns3::DroneList}.

According to the peculiar workflow of \ac{ns-3}, to properly instantiate a \texttt{ns3::Drone} object, a \texttt{ns3::DroneContainer} is needed. The creation process consists in a call to the \texttt{ns3::Object::CreateObject<T>} function, where \texttt{T} is replaced with \texttt{Drone}. In order to ensure full compatibility with all \ac{ns-3} methods involving \texttt{ns3::Node} or \texttt{ns3::NodeContainer} classes, a dedicated mechanism has been developed. Every \texttt{ns3::Drone} goes through a static cast procedure, i.e. \texttt{ns3::StaticCast}, which generates a \texttt{ns3::Node} object that is pushed into a \texttt{ns3::NodeContainer}. In this way, for each drone, two smart pointers refer to the same memory location, but cast to the two required types.
Besides, the \texttt{ns3::DroneContainer} class provides a specific iterator, together with two further methods which return the number of instantiated drones, and a smart pointer to each.
It is worth mentioning that only drones must use a \texttt{ns3::DroneContainer}, while \acp{ZSP}, together with other entities, must still be modeled as \texttt{ns3::Node} objects.

\subsubsection{Peripherals}
A \ac{UAV} is usually equipped with a set of peripherals able to extend its capabilities. Such peripherals include a wide range of devices, implemented in \ac{IoD-Sim}, through new specific classes.
The \texttt{ns3::DronePeripheral} object represents a general-purpose on-board peripheral with the following properties:
\begin{itemize}
    \item \textit{Peripheral state} -- which can either be set to \texttt{ON}, \texttt{OFF}, or \texttt{IDLE}. This simple \ac{FSM} allows the development of intelligent algorithms to find optimal energy management.
    \item \textit{Power consumption} -- how much instantaneous power is required by the peripheral, expressed in Watts, for each state.
    \item \textit{Reference \acp{RoI}} -- where the peripheral should be operating. This is extremely useful to model certain peripherals and missions that depend on particular regions in space. For instance, a photo camera can be used and activated only when the drone is in the \ac{RoI}, thus leading to an optimized use of power, storage, and data. If this parameter is not defined, the reference peripheral will be active over time.
\end{itemize}
\texttt{ns3::DronePeripheral} has been specialized in two subclasses.

\texttt{ns3::StoragePeripheral} represents a generic storage device characterized by an attribute describing the initial amount of memory, which can be traced at runtime to record the empty space left. Device total capacity can be queried through \texttt{GetCapacity()} method. If a drone peripheral, e.g. camera or any other sensor, wants to interact with the storage, it is possible to request space by specifying the amount of data through \texttt{Alloc()}. The inverse can be done with \texttt{Free()}. These operations can fail if there is no memory left or there are no data to be freed, respectively. For this reason, a boolean value is returned by these methods to indicate if the requested operation was successfully completed or not.
In this work, it is assumed that at most one \texttt{ns3::StoragePeripheral} is installed on each drone.

\texttt{ns3::InputPeripheral} describes a generic input device, characterized by an acquisition \texttt{DataRate}, constant over a \texttt{DataAcquisitionTimeInterval}.
Once it is created, installed on a drone, and attached on a particular storage peripheral with \texttt{Install()} method, the storage peripheral of reference can be changed with \texttt{SetStorage()}. If the peripheral is \texttt{ON}, \texttt{AcquireData()} simulates data acquisition at the given \texttt{DataRate}.

These two peripheral types are strongly connected, since a \texttt{ns3::InputPeripheral} can offload acquired data to a \texttt{ns3::StoragePeripheral} through a boolean attribute. Nonetheless, the association between input and storage is not mandatory. In fact, in a real-world scenario, an \texttt{ns3::InputPeripheral} can deliver data directly to a processing unit or to a remote host, thus neglecting the need to permanently store the information.

A complete list of the attributes of these classes is given with Table \ref{tab:drone_per_attr}. It is worth specifying that all peripherals hold a reference to the drone they are equipped to.

\begin{table*}
    \caption{Drone Peripherals Properties.}
    \label{tab:drone_per_attr}
    \centering
    \begin{tabular}{llr}
        \toprule
        Class & Attribute & Description \\
        \midrule
        \texttt{DronePeripheral} & \texttt{PowerConsumption} & Power consumption of the peripheral in J/s \\

        \texttt{StoragePeripheral} & \texttt{Capacity} & The capacity of the disk in bit \\
        & \texttt{DataRate} & The acquisition data rate of the peripheral in bit \\

        \multirow{3}*{\texttt{InputPeripheral}} & \texttt{InitialRemainingCapacity} & The starting remaining capacity in bit \\
        & \texttt{DataAcquisitionTimeInterval} & The time interval occurring between any data acquisition \\
        & \texttt{HasStorage} & Acquired data are offloaded to the StoragePeripheral \\
        \bottomrule
    \end{tabular}
\end{table*}

Moreover, for each \texttt{ns3::Drone}, a \texttt{ns3::DronePeripheralContainer} object is created to manage all its peripherals. This container is responsible for the creation of peripherals and, through the \texttt{ns3::DronePeripheralContainer::InstallAll()} method, finalizes the initialization, sets the correct references to the host drone, and, eventually, to the target \texttt{ns3::StoragePeripheral}.

\subsubsection{Mechanics and Energy Consumption}
\ac{ns-3} already models and manage all the energy-related aspects, such as consumption, harvesting, and monitoring, through the abstract class \texttt{ns3::EnergySource}. Although there is no specific energy source model available that is suitable for drones, the \texttt{ns3::LiIonEnergySource} is sufficiently general to be employed for simulation purposes \cite{shepherd1963theoretical,tremblay2007generic}.

The \texttt{ns3::DeviceEnergyModel} class describes the \texttt{ns3::NetDevice} energy consumption, by means of the drawn current. The installation procedure is eased by the helper class \texttt{ns3::DeviceEnergyModelHelper}, which employs the \texttt{Install()} method, that links a \texttt{ns3::EnergySource} to a \texttt{ns3::NetDevice}.

When the battery object is initialized, it schedules an \texttt{ns3::Event}, which calls \texttt{ns3::EnergySource::CalculateTotalCurrent()}. This function retrieves the current drawn of every device associated with the \texttt{ns3::EnergySource}, by calling \texttt{ns3::DeviceEnergyModel::GetCurrentA()}. Subsequently, the energy consumption value is calculated and subtracted from the remaining one. Finally, the \texttt{ns3::Event} reschedules itself.

In this work a specialization of \texttt{ns3::DeviceEnergyModel}, i.e. \texttt{ns3::DroneEnergyModel}, is developed along with the helper class \texttt{ns3::DroneEnergyModelHelper}. Given a simulation duration $T$, the model splits it into $n=1,\ldots,N$ equal discrete intervals. The power consumption model of the drone flying at speed $\textbf{v}[n] = (v_x[n], v_y[n], v_z[n])$, in the $n$-th time slot, is the following \cite{SXN+19}:
\begin{equation}
    P_{UAV}[n]=P_{level}[n]+ P_{vertical}[n]+ P_{drag}[n],
\end{equation}
where
\begin{align}
    P_{level}[n] &= \frac{W^2}{\sqrt{2}\rho A} \frac{1}{\sqrt{\Omega+\sqrt{\Omega^2+4V^4_h}}},
\end{align}
being
\begin{align}
    \Omega &= {\Vert(v_x[n],v_y[n])\Vert}^2\\
    P_{vertical}[n]&=Wv_z[n],\\
    P_{drag}[n]&=\frac{1}{8}C_{D0}\rho A {\Vert(v_x[n],v_y[n])\Vert}^3,
\end{align}
$W = mg$, with $m$ defining the mass of the drone and $g$ as the gravitational acceleration. Moreover, $\rho$ is the air density, $A$ is the total rotor disk area,
$C_{D0}$ is the profile drag coefficient depending on the geometry of the rotor blades, and $V_h = \sqrt{\frac{W}{2\rho A}}$ uses parameters to calculate the power required for hovering operations.

The energy model can be aggregated to a drone by means of the \texttt{ns3::Drone} \texttt{EnergyModelHelper}, which provides an \texttt{Install()} method that aggregates it to \texttt{ns3::Drone}. In this way, it is possible to simulate the energy characteristics of a drone, both for its mechanics and its peripherals, in addition to its networking operations.

Such mechanical power consumption model is implemented in the method \texttt{ns3::DroneEnergyModel::GetPower()}. Similarly, the method \texttt{ns3::DroneEnergyModel} \texttt{::GetPeripheralsPowerConsumption()} returns the cumulative power consumption of all peripherals on board.

The \texttt{ns3::DroneEnergyModel} object, registered as a new \texttt{ns3::TypeId} with no attributes, implements \texttt{ns3::DoGetCurrentA()} inherited from \texttt{ns3::DeviceEnergyModel}.
Such method returns the total drawn current related to both mechanics and peripherals, in addition to networking operations.
The energy model can be aggregated to a drone by means of \texttt{DroneEnergyModelHelper}, which provides an \texttt{Install()} method that aggregates it to \texttt{ns3::Drone}.

It is worth specifying that \texttt{ns3::DroneEnergyModelHelper} implements the installation procedure in a different manner with respect to its parent, i.e., \texttt{ns3::DeviceEnergyModelHelper}. In fact, the \texttt{ns3::DroneEnergyModelHelper::Install()} method links a \texttt{ns3::EnergySource} to a \texttt{ns3::Drone}, instead of a \texttt{ns3::NetDevice}.
This aspect distinguishes the aim of \ac{IoD-Sim} from the \ac{ns-3} one: to simulate all the relevant aspects of the drone, beyond networking perspective. This justifies the implementation divergence from the \ac{ns-3} main goals.

During the simulation, it is possible that the drone goes out of energy. To this end, the event is propagated through the execution of \texttt{HandleEnergyDepletion()} of the energy model, for which the time of depletion is logged for successive data analysis.

\subsection{Other Simulation Entities: \acsp{ZSP} and Remotes}
Entities beyond \texttt{ns3::Drone} are \textit{\acp{ZSP}} and \textit{Remotes}. \acp{ZSP} are smart entities, modeled as \texttt{ns3::Node} objects, equipped with multiple \texttt{ns3::NetDevice} which provide multi-protocol radio access, thus enabling communications between drones and the rest of the Internet. Typically, they are configured as ground entities that maintain a constant position in time \cite{GBW16}, by means of \texttt{ns3::ConstantPositionMobilityModel}. Nonetheless, in \ac{IoD-Sim} their mobility model can be customized to fit simulation purposes, envisioning the adoption of dynamic wireless infrastructure proposed in 5G \& Beyond architectures. \textit{Remotes}, instead, are \texttt{ns3::Node} objects with no mobility model and only rely on installed applications which provide remote services to consumers. Remotes and \acp{ZSP} are interconnected through a backbone, simplified as a \ac{CSMA}-based bus network, that represents the \textit{Internet}. This architecture allows service provisioning on different classes of nodes, employing \textit{Remotes} in case of applications with high computational costs, e.g., multimedia data processing, and \acp{ZSP} in case of low latency requirements, e.g., traffic management.

\subsection{Mobility}
\ac{ns-3} provides a basic foundation to represent the movement of drones (e.g., \verb|ns3::WaypointMobilityModel|, \verb|ns3::ConstantAccelerationMobilityModel|, and \verb|ns3::ConstantVelocityMobilityModel|). However, an important gap arises when such models are analysed in details: none of the available ones are able to construct a curve trajectory that take into account how much a spot is relevant for the mission plan.
Another aspect to consider is that models such as \verb|ns3::WaypointMobilityModel| couples the position of the drone with a given time instant, without taking into account the limitations imposed by the maximum speed of the \ac{UAV}. Therefore, if the user does not properly design the path, this could lead to a simulation which does not reflect the reality. Moreover, in the setup phase it is necessary to specify all the points that create the trajectory.

To overcome these limitations, dedicated mobility models have been developed. In particular, the trajectory has been modeled using B\'ezier curves by specifying a set of \ac{PpoI}. These are decoupled from the time of arrival, and the resulting trajectory is bounded to the mechanical characteristics of the drone.
A specific structure implemented in \ac{IoD-Sim}, namely \verb|ns3::CurvePoint|, describes the 3D position vector of the B\'ezier curve together with the distances from the previous point and the starting one.
Besides, a container object, i.e., \verb|ns3::Curve|, is in charge of managing the points of the curve, i.e., \verb|ns3::CurvePoint|, that are defined according to the interest points contained in a \verb|ns3::FlightPlan|.
When a \verb|ns3::Curve| is instantiated, it populates the container according to the following.

Let $\textbf{P} = \bigl\{ \textbf{P}_0, \textbf{P}_1, \dots, \textbf{P}_{N-1} \bigr\}$ with $\textbf{P}_i \in \mathbb{R}^3, \ \forall i=0,\ldots,N-1$ be an ordered sequence of $N$ interest points, $\textbf{l} = \bigl\{ l_0, l_1, \dots, l_{N-1} \bigr\}$, $l_i \in \mathbb{N}^+$, the interest level associated to each point, $\Lambda = \left( \sum_{i=0}^{N-1} l_i \right) - 1$ and $L_i = \sum_{h=0}^{i-1} l_h$.
The \textit{Trajectory Generator} can be expressed as
\begin{equation}
    \textbf{G}(t) = \sum_{i=0}^{N-1} \textbf{P}_i \sum_{j=0}^{l_i - 1} \binom{\Lambda}{L_i + j} (1-t)^{\Lambda-L_i-j} t^{L_i+j}, \quad t \in [0, 1]
    \label{eq:trajectoryGenerator}
\end{equation}

\begin{figure}
    \centering
    \includegraphics[width=\columnwidth]{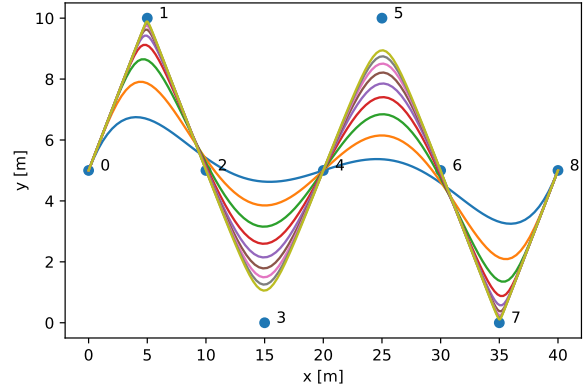}
    \caption{A set of trajectories, generated with \eqref{eq:trajectoryGenerator}, with different Interest Levels (from 1 to 10, incrementally) for \ac{PpoI} 1, 3, 5, and 7. The other points have constant Interest Level set to 1.}
    \label{fig:trajectoryGenerator}
\end{figure}

It is worth noting that \eqref{eq:trajectoryGenerator} is a revised version of the original B\'ezier equation, which does not practically allow to reach the interest points, except for the first and last one.
An increment in the interest level $l$ turns into a trajectory that passes closer to that point, as illustrated in Figure \ref{fig:trajectoryGenerator}.
A special case takes place when $l = 0$. A specific mechanism is provided to split the trajectory into two contiguous curves, so that the drone is forced to fly over them. In this case, a \verb|restTime| can be defined to set the hovering duration in seconds.

Finally, the obtained trajectory is used by the new implemented models, i.e., \verb|ns3::ConstantAccelerationDroneMobilityModel| and
\verb|ns3::ParametricSpeedDroneMobilityModel|.

\subsubsection{Constant Acceleration Drone Mobility Model}
This mobility model employs \eqref{eq:trajectoryGenerator} and the uniform acceleration motion law to retrieve the points of the desired trajectory. Since the speed of the drone cannot increase indefinitely, after the maximum speed is reached, the uniform linear motion law is adopted.

This object is implemented as \ac{ns-3} model, and hence, has its own \texttt{TypeId} with attributes described in Table \ref{tab:attributiContantAccelerationDroneMobilityModel}.

\begin{table}[htbp]
    \caption{\texttt{ns3::ConstantAccelerationDrone MobilityModel} TypeId attributes.}
    \label{tab:attributiContantAccelerationDroneMobilityModel}
    \centering
    \begin{tabular}{lp{5cm}}
        \toprule
        Attribute & Description \\
        \midrule
        \texttt{Acceleration} & Drone's constant acceleration, expressed in \si{m/s^2}.  \\
        \texttt{MaxSpeed} & Drone's maximum speed, expressed in \si{m/s}. \\
        \texttt{FlightPlan} & Interest points for the trajectory. \\
        \texttt{SimulationDuration} & Simulation duration, expressed in seconds. \\
        \texttt{CurveStep} & Discretization step of the curve. \\
        \bottomrule
    \end{tabular}
\end{table}
\begin{table}[htbp]
    \caption{\texttt{ns3::ParametricSpeedDrone MobilityModel} TypeId attributes.}
    \label{tab:attributiParametricSpeedDroneMobilityModel}
    \centering
    \begin{tabular}{lp{5cm}}
        \toprule
        Attribute & Description \\
        \midrule
        \texttt{SpeedCoefficients} & The set of coefficients for the polynomial $v(t)$. \\
        \texttt{FlightPlan} & Interest points of the trajectory. \\
        \texttt{SimulationDuration} & Simulation duration, expressed in seconds. \\
        \texttt{CurveStep} & Discretization step of the curve. \\
        \bottomrule
    \end{tabular}
\end{table}
\begin{table*}[!ht]
    \caption{Configuration parameters for \textit{Telemetry Applications}.}
    \label{tab:droneApplicationsParams}
    \centering
    \begin{tabular}{llp{2cm}lp{5cm}}
        \toprule
        Application Type & Name & Type & Default Value & Description \\
        \midrule
        Client & \verb|DestinationIpv4Address| & String & \verb|255.255.255.255| & IPv4 address of the remote application server. \\
        Client and Server & \verb|Port| & Unsigned Integer 32-bit & \verb|80| & Port of the remote application server or listening port in case of the server. \\
        Client & \verb|TransmissionInterval| & Double & \verb|1.0| & Transmission interval of the telemetry updates being sent, in seconds. \\
        Client and Server & \verb|StartTime| & Double & Start of Simulation & Time at which to start the application, in seconds. \\
        Client and Server & \verb|StopTime| & Double & End of Simulation & Time at which to stop the application, in seconds. \\
        Client & \verb|FreeData| & Boolean & \verb|false| & Free data from the equipped storage peripheral when they are transmitted. \\
        Server & \verb|StoreData| & Boolean & \verb|false| & Store data to the equipped storage peripheral when they are received. \\
        \bottomrule
    \end{tabular}
\end{table*}

In each instant of the simulation, \ac{IoD-Sim} calls two methods, \verb|DoGetPosition ()| and \verb|DoGetVelocity ()|. They return both the position and the speed at current time of the drone, that is recomputed thanks to the \verb|Update ()| method.

\subsubsection{Parametric Speed Drone Mobility Model}
Similarly to \textit{Constant Acceleration Drone Mobility Model}, also this mobility model is implemented as a \ac{ns-3} model with its own \verb|TypeId|. However, this takes a $v(t)$ speed profile in a polynomial form and, thanks to the modified B\'ezier equation \eqref{eq:trajectoryGenerator}, it retrieves the discretized trajectory.
To ease the implementation, a specific attribute, i.e., \verb|ns3::SpeedCoefficients|, is introduced to serve as a container of the $v(t)$ coefficients.
These are elaborated (by employing the \ac{GSL}) to constantly update the parameters by calling \verb|UpdateSpeed ()| and \verb|UpdatePosition ()| subroutines.
A summary of the attributes of this mobility model is reported in Table \ref{tab:attributiParametricSpeedDroneMobilityModel}.

\subsection{Applications}
\ac{IoD-Sim} offers simple applications that can be used to communicate telemetry from a drone to a \ac{ZSP} or to a \textit{Remote} by adopting client-server paradigm, via \ac{UDP}.
Moreover, relying on the same architecture, two \ac{TCP}-based applications are available to enable reliable data transfer between hosts.
Besides, a \ac{NAT}-like application is provided to design relaying network architectures.

\subsubsection{Telemetry Applications}\label{sec:telemetry_applications}
These applications are modeled as classes named \verb|ns3::DroneClientApplication| and \verb|ns3::DroneServerApplication|. The model asks for the \texttt{DestinationIpv4Address} and a \texttt{Port} of the remote entity that hosts the server application. Data are sent every \texttt{TransmissionInterval} and, whereas the drone has a storage peripheral, it is possible to free an equivalent amount of memory space. The configuration parameters are summarized in Table \ref{tab:droneApplicationsParams}.

\begin{table*}[ht]
    \caption{Configuration parameters for \textit{Generic Traffic Applications}.}
    \label{tab:genericTrafficAppsParams}
    \centering
    \begin{tabular}{lp{4cm}p{2cm}lp{5cm}}
        \toprule
        Application Type & Name & Type & Default Value & Description \\
        \midrule
        All Server and Clients & \texttt{Address} & String & \texttt{127.0.0.1} & Listening or remote address of the server. \\
        All Server and Clients & \texttt{Port} & Unsigned Integer 16-bit & \texttt{4242} & Listening or remote port of the server. \\
        All Clients & \texttt{PayloadSize} & Unsigned Integer 16-bit & \texttt{65470} & Size of the payload for each packet, in bytes. In case of Storage Client, it is the maximum size to be used when freeing storage memory. \\
        Periodic Client only & \texttt{Frequency} & Double & \texttt{1.0} & Number of times in a second when a new packet is sent to the server. \\
        \bottomrule
    \end{tabular}
\end{table*}

\begin{figure}[ht]
    \centering
    \includegraphics[width=\columnwidth]{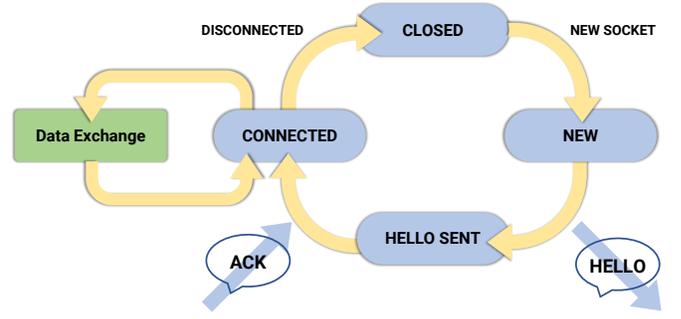}
    \caption{\ac{FSM} of the Drone Client and Server Application.}
    \label{fig:applicationstates}
\end{figure}

When the application is started, through the \verb|ns3::Application::StartApplication()| method, a \ac{UDP}-based communication, employing application-level acknowledgements, takes place. It is worth specifying that the application is stateful in order to support the \textit{Rendezvous Process} which discovers the application server in the network, if no address is given. This process starts with the client application in \verb|NEW| state. Therefore, a \verb|HELLO| packet is sent to the destination address (or in broadcast), thus implying a state transition in \verb|HELLO_SENT|. If the application server receives such packet, it replies with an \verb|HELLO_ACK| packet to confirm the reception. When the client receives the acknowledgement, its state changes again, into \verb|CONNECTED|, which allows it to periodically send telemetry data. These packets are named \verb|UPDATE| and \verb|UPDATE_ACK|. The entire procedure is depicted in Figure \ref{fig:applicationstates}.

The \ac{JSON}-encoded telemetry is periodically transmitted, through the \verb|SendPacket()| method, and received by the application server, through the \verb|ReceivePacket()| method. \texttt{HELLO} and \texttt{UPDATE} packets transport a payload which is formatted in \ac{JSON} with ASCII encoding. Its content is a \ac{JSON} object with the following properties:
\begin{itemize}
    \item The unique \texttt{id} of the drone in the simulation. This ensures that \textit{Drones} communications can be tracked over complex scenarios.
    \item An incremental \texttt{sn} that refers to the \textit{sequence number}. It is used to easily check if a packet has been lost.
    \item \texttt{cmd} that refers to the type of packet, whether if \texttt{HELLO}, \texttt{UPDATE}, or an acknowledgment.
    \item \texttt{gps} coordinates with \texttt{lat} for latitude, \texttt{lon} for longitude, \texttt{alt} for altitude, and \texttt{vel} for the velocity vector. For simulated drones, the GPS location refers to the virtual world coordinates.
\end{itemize}
The \ac{UDP} packet payload is summarized in Table \ref{tab:udppayload}. When the application is stopped, the \verb|StopApplication()| method is called.
\begin{table}[ht]
    \caption{UDP payload.}
    \label{tab:udppayload}
    \centering
    \begin{tabular}{p{.3cm}p{.3cm}ll}
        \toprule
        \multicolumn{2}{l}{Field Name} & Data Type & Description \\
        \midrule
        \verb|id| && Unsigned Integer 32-bit & \ac{ns-3} Global Node Identifier \\
        \verb|sn| && Unsigned Integer 32-bit & Packet Sequence Number \\
        \verb|cmd| && String & Packet Type \\
        \verb|gps| && Object & Drone location in space \\
        & \verb|lat| & Double & Drone latitude \\
        & \verb|lon| & Double & Drone longitude \\
        & \verb|alt| & Double & Drone altitude \\
        \verb|vel| && Array of 32-bit Integers & Drone velocity in m/s \\
        \bottomrule
    \end{tabular}
\end{table}

Clearly, these applications are developed so that multiple instances can run concurrently, on the same entity, if different ports are specified. Moreover, they are independent from the particular communication technology adopted.

\subsubsection{Generic Traffic Applications}\label{sec:generic_traffic_applications}
These applications model a reliable data transfer between a client and a server, which are implemented as \texttt{TcpPeriodicClientApplication} and \texttt{TcpEchoServerApplication} objects, respectively. The aim is to transfer a certain amount of information between the two hosts according to the specified \texttt{PayloadSize}, expressed in bytes, and \texttt{TransmissionFrequency}, measured in Hz, set on the client. The server is characterized by a socket, composed by a listening \texttt{Address} and \texttt{Port}. These configuration parameters are summarized in Table \ref{tab:genericTrafficAppsParams}.
To facilitate traffic analysis, each packet has a \ac{PDU}, formed by a 12 bytes header, and the payload. The former contains information-level sequence number and the timestamp of creation; the latter is characterized by a recurring sequence of a 16 bit number that is incremented over time.
These applications provide dedicated \texttt{TraceSource} objects that notify communication-related events such as new/closed connections and sent/received packets.

An additional \ac{TCP}-based client has been create to support drones that are typically equipped with a \texttt{StoragePeripheral}. To this end, \texttt{TcpStorageClientApplication} monitors the storage and, if memory is used, it transfers data to the remote server. If the transfer is acknowledged, memory is freed. This mechanism is of relevance when drones are equipped with limited on-board memory. Indeed, the client can be used to transfer as much data as possible over the wireless medium to prevent out-of-memory events.

\subsubsection{Relaying Application}\label{sec:relayapp}
The \textit{Relaying Application} is implemented through the class \texttt{ns3::NatApplication}. It is a specialized networking application that, given an \texttt{InternalNetDeviceId} and an \texttt{ExternalNetDeviceId}, provides a \ac{NAT}-like mechanism to a set of drones placed in an internal network.
The \textit{NetDeviceId} is a numerical identifier that uniquely points to a network device mounted on the drone.

During initialization, i.e., \texttt{DoInitialize()} method, the application modifies the static routing table of the internal network device to redirect all traffic to the loopback device. A specific callback, namely \texttt{RecvPktFromNetDev()}, notifies when a new frame arrives.
It contains information such as the \ac{IANA} standard L3 protocol identifier and the sender/receiver MAC addresses.

The \ac{NAT} forwarding behavior leverages a hash map, i.e., \textit{\ac{NAT} Table}, where an external port number is coupled with the source IP address and port. Inbound frames are forwarded to the external network by replacing this information with the one of the relaying drone. The same rationale is applied for frames received from the external network.

\subsection{Scenario Configuration Interface}
\begin{table}[ht]
    \caption{Memory organization of protocol stacks in the General Purpose Scenario.}
    \label{fig:protocolStack}
    \centering
    \begin{tabular}{c||cccc}
        \\ \hline
        \backslashbox{\textbf{Stack ID}}{\textbf{Layer}} & \textbf{PHY} & \textbf{MAC} & \textbf{NET} \\
        \hline
        0 & \verb|WifiPhy| & \verb|WifiMac| & \verb|IPv4| \\
        1 & \verb|WifiPhy| & \verb|WifiMac| & \verb|IPv6| \\
        2 & \verb|LtePhy| & \verb|LteMac| & \verb|IPv4| \\
        \vdots & \vdots & \vdots & \vdots \\
        \hline
    \end{tabular}
    \tikz[remember picture, overlay]{%
      \node at (-4,1.5) {\texttt{std::array}};
      \node (A) at (0,1.5) {};
      \node (B) at (-3,1.5) {};
      \node (C) at (-5,1.5) {};
      \node (D) at (-7.6,1.5) {};
      \draw[-{Stealth[width=10]},draw opacity=0] (B) edge (A);
      \draw[-{Stealth[width=10]},draw opacity=0] (C) edge (D);
      \node[rotate=90] at (-7.83,0.1) {\texttt{std::vector}};
      \node (E) at (-7.8,1.4) {};
      \node (F) at (-7.8,-1.2) {};
      \draw[-{Stealth[width=10]},draw opacity=0] (E) edge (F);
      \draw[-{Stealth[width=10]},draw opacity=0] (F) edge (E);
      }
\end{table}

The \textit{Scenario Configuration Interface} is an abstraction layer that allows the configuration of the entire simulation by means of \ac{JSON} files.
Indeed, they can be decoded and validated through RapidJSON in order to setup the simulation models.
The output data classes are then used by the \textit{General Purpose Scenario} to initialize objects that define the environment, the entities, and the simulator engine.
To this end, the set of all objects that are used to characterize a scenario can be grouped into three categories:
\begin{itemize}
    \item \textit{Configuration Objects} -- Models that store parameters in a structured way, easily accessible in the C++ language.
    \item \textit{Configuration Helpers} -- Checkers and decoders with the goal to produce a Configuration object or throw an error message.
    \item \textit{Simulation Helpers} -- Objects that help organise pointers to structures commonly found in scenario development. They are used in the protocol stack matrix, shown in Table \ref{fig:protocolStack}.
\end{itemize}
Additionally, \textit{Factory Helpers} are defined as weakly-coupled extensions to \ac{ns-3} internal data structures to ease their initialisation. They are made to minimise modifications done to the \ac{ns-3} core framework, used by \ac{IoD-Sim}.
The entire system has been made extensible by design, so that it is possible to support further technologies and configurations with the addition of new configuration objects and helpers as needed.
In this way it is possible to further develop high-level configuration interfaces able to setup scenarios and hence easing the design activity made by the user.

\subsubsection{Scenario Configuration Objects and Helpers}
The core of the abstraction layer is the \texttt{ns3::ScenarioConfigurationHelper}, a low-level object that directly deals with the \ac{JSON} configuration file. This helper returns a set of specific data classes that contain exclusively the parameters required to configure \ac{IoD-Sim} models. Each of them is also loosely coupled with a \ac{JSON} validator and parser, also known as configuration helper. The information embedded in these classes is then deserialized and employed by higher-level objects.

\begin{itemize}
    \item \texttt{ns3::ModelConfiguration} describes \texttt{ns3::TypeId} objects through key-value pairs that reference the model attributes.
    \item \texttt{ns3::EntityConfiguration} describes an entity, whether it is a \textit{Drone}, a \ac{ZSP}, or a \textit{Remote}. The object retrieves and stores all parameters related to the \texttt{ns3::NetDevice} to be installed on the entity, the \textit{Mobility Model} to be applied, and the \textit{Applications}. Optionally, if the entity is a \textit{Drone} there can be defined the mechanics, the battery, and the peripherals. Its parser is called \texttt{ns3::EntityConfigurationHelper}.
    \item \texttt{ns3::RemoteConfiguration} denotes key characteristics of \textit{Remotes}. Specifically, a remote need to know the global network layer ID of reference and the configurations of applications to be installed. Its parser is \texttt{ns3::RemoteConfigurationHelper}.
    \item \texttt{ns3::PhyLayerConfiguration} defines the required parameters needed to configure a PHY layer. It is the parent and interface of \texttt{ns3::LtePhyLayerConfiguration} and \texttt{ns3::WifiPhyLayerConfiguration} data classes. Its parser is \texttt{ns3::PhyLayerConfigurationHelper}.
    \item The \texttt{ns3::LtePhyLayerConfiguration} gets all the information needed to set up a PHY layer for \ac{LTE}, such as its propagation loss model and its spectrum model.
    \item \texttt{ns3::WifiPhyLayerConfiguration} sets up the PHY layer of a Wi-Fi based protocol stack. The PHY layer configuration requires the higher-level Wi-Fi standard to be used, the antenna Rx gain, the data rate, the propagation delay and loss models.
    \item \texttt{ns3::MacLayerConfiguration} collects the required parameters needed to configure a MAC Layer. It is the parent and interface of \texttt{ns3::WifiMacLayerConfiguration}.  Its parser is \texttt{ns3::MacLayerConfigurationHelper}.
    \item \texttt{ns3::WifiMacLayerConfiguration} configures a Wi-Fi \ac{BSS}. The \ac{SSID} and access point parameters are defined to create its basic infrastructure.
    \item \texttt{ns3::NetworkLayerConfiguration} defines the required parameter needed to configure appropriately a network layer. It is parent to the \texttt{ns3::Ipv4NetworkLayerConfiguration}. Its parser is named \texttt{ns3::NetworkConfiguration- Helper}.
    \item \texttt{ns3::Ipv4NetworkLayerConfiguration} stores the network address and mask of the configured IPv4 Layer in the configuration file.
    \item \texttt{ns3::LteBearerConfiguration} decodes all the relevant parameters for a \ac{LTE} bearer, such as its type and the \ac{QoS} defined as a tuple of \textit{Guaranteed Bit Rate} and \textit{Maximum Bit Rate}.
    \item \texttt{ns3::LteNetdeviceConfiguration} collects the information needed by an \ac{LTE} network device, such as its bearers. The \textit{role} of the network device is then detected, whether it is a User Equipment or an eNB.
    \item \texttt{ns3::NetdeviceConfiguration} defines for a generic network device. The main parameter stored is the global network layer ID, which is used to detect the stack and network to be attached when the network device is created and installed on a \textit{Node}. A specific configuration for Wi-Fi network devices is handled by \texttt{ns3::WifiNetdeviceConfiguration} with relevant MAC data to connect to the \ac{BSS}.
\end{itemize}

\subsubsection{Scenario Simulation Helpers}
To enable complex scenarios that are related to the future \ac{IoD} communication paradigms, \ac{IoD-Sim} enables the simulation of \ac{IoD} networks in which multiple telecommunication protocols are used at the same time, both for the drones and the \acp{ZSP}.
Currently, \ac{IoD-Sim} supports two communication technologies that can be used concurrently: \ac{LTE} and the IEEE 802.11 family. Each protocol stack must be applied to a dedicated network device, i.e., \verb|ns3::NetDevice|.
The architecture established in the simulator has been designed so that it eases the configuration phase.

In order to facilitate the implementation and the installation of protocol stacks on \ac{IoD} entities, additional helpers named \textit{Simulation Helpers} have been developed to arrange the necessary common infrastructure to simulate communications among nodes.
Thus, the developed \textit{Simulation Helpers} are:
\begin{itemize}
    \item \verb|ns3::WifiPhySimulationHelper|, that initializes the PHY layer of a Wi-Fi-based protocol stack.
    \item \verb|ns3::WifiMacSimulationHelper|, that creates the objects related to IEEE 802.11 MAC.
    \item \verb|ns3::LtePhySimulationHelper|, that allocates the necessary resources to enable \ac{LTE} communications.
    \item \verb|ns3::Ipv4SimulationHelper|, that manages IPv4 networks for each protocol stack.
\end{itemize}
All the aforementioned can cooperate with the existing helpers in \ac{ns-3}, such as \verb|ns3::LteHelper|, \verb|ns3::WifiHelper|, \verb|ns3::YansWifiPhyHelper|, and \verb|ns3::WifiMacHelper|.

\subsubsection{General Purpose Scenario}
\begin{figure}
    \centering
    \includegraphics[width=0.9\columnwidth]{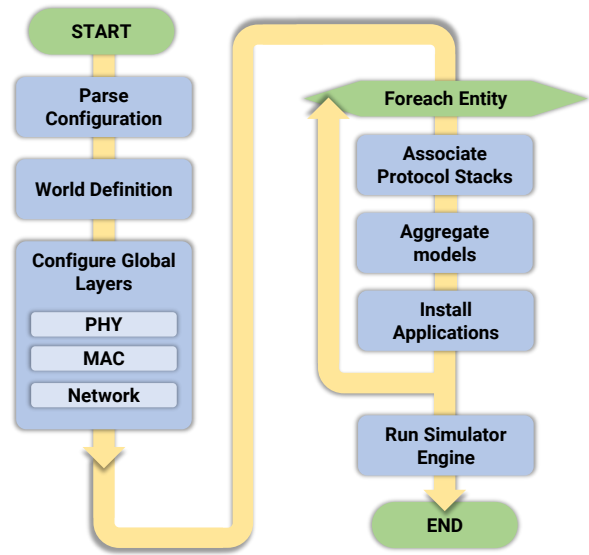}
    \caption{Logical flow to initialise and configure a scenario in \acs{IoD} Sim.}
    \label{fig:flowchart}
\end{figure}

A flexible and highly dynamic \textit{General Purpose Scenario} has been developed in order to setup scenario's entities and, at the same time, to provide abstractions which minimize the effort from a programming perspective. It is fully dependent on a semantic analyzer and allows the entire simulation platform to be compiled beforehand, providing ways to dynamically reconfigure the scenario at run-time.
Its development started from the analysis and the detection of a common structure typically followed by the \ac{OSI} protocol stack. The entire workflow, depicted in Figure \ref{fig:flowchart}, is described hereby.

\textit{General Purpose Scenario} is composed in two main parts: \textit{configuration} and \textit{run}. Scenario \textit{configuration}, executed through the constructor \texttt{Scenario()}, is interleaved with the \textit{Scenario Configuration Interface}. The \textit{run} part is identified by \texttt{operator()()} which is characterized by minimal C++ code that starts the \ac{ns-3} simulator engine. Moreover, it shows the progress status on the console and, optionally, it saves messages on a log file.

The \textit{General Purpose Scenario} requires the initialization of the \textit{Scenario Configuration Interface} through \ac{JSON} configuration file. Once the file is decoded, the number of entities are retrieved to create the initial structures, such as a \texttt{ns3::DroneContainer} and four \texttt{ns3::NodeContainer} objects. They keep track of \acp{ZSP}, \textit{Remotes}, and nodes that participate on the \textit{Backbone Network}.

Once the entities are created, they are registered to their respective global lists, such as \texttt{ns3::DroneList}, \texttt{ns3::ZspList}, and \texttt{ns3::RemoteList}.

After entity creation, the \ac{ns-3} static configuration parameters are applied to the simulation. The method is called \texttt{ApplyStaticConfig()}. These parameters are a set of key-value pairs that represent certain features of \ac{ns-3} models.

World definition is made through \texttt{ConfigureWorld()} method. It is related to the configuration of \textit{Buildings} and \acp{RoI}.
The virtual world set up is then followed by the configuration of PHY, MAC, and Network global layers.

As for the PHY layer part, if it is made for a Wi-Fi communication stack, the \texttt{ns3::WifiPhySimulationHelper} is employed with the specifications stored in \texttt{ns3::WifiPhyLayerConfiguration}. If the PHY layer is for \ac{LTE}, instead, the \texttt{ns3::LtePhySimulationHelper} is set up with \texttt{ns3::LtePhyLayerConfiguration} parameters. The same procedure is applied for the global MAC layer configuration.
The global \textit{Network} layer is managed by \texttt{ns3::Ipv4SimulationHelper} for IPv4 networks with the specifications given by \texttt{ns3::NetworkLayerConfiguration}, i.e., network address, mask, and a default route.

Global stacks are then linked to the configured entities. Moreover, for \ac{LTE} devices, the bearer is created to ensure that applications have a logical communication channel with desired properties.
When the entity network configuration is done, the mobility model is configured and the applications are installed. Furthermore, if the entity is a \textit{Drone}, its peripherals are installed, together with the associated energy model.

Once all entities are ready, the virtual internet backbone is configured. A \ac{CSMA} bus is made for the backbone network, identified with address \texttt{200.0.0.0/8}. Hosts that can be part of this backbone network are \textit{Remotes}, \textit{Packet Gateways} in case of \ac{LTE} core network, or other routers in case of the presence of a Wi-Fi \ac{BSS}.

Finally, in case of \ac{LTE} networks, their Radio Environment Maps are set up to generate images that represent the radiation map of the \ac{RAN}.

\subsubsection{JSON Configuration Schema}\label{sec:jsonconf}
The entire scenario has been made parametric through the use of a \ac{JSON} configuration file. Requested at startup, it is decoded and employed to configure and execute the simulation.

\begin{figure}
    \centering
    \includegraphics[width=\columnwidth]{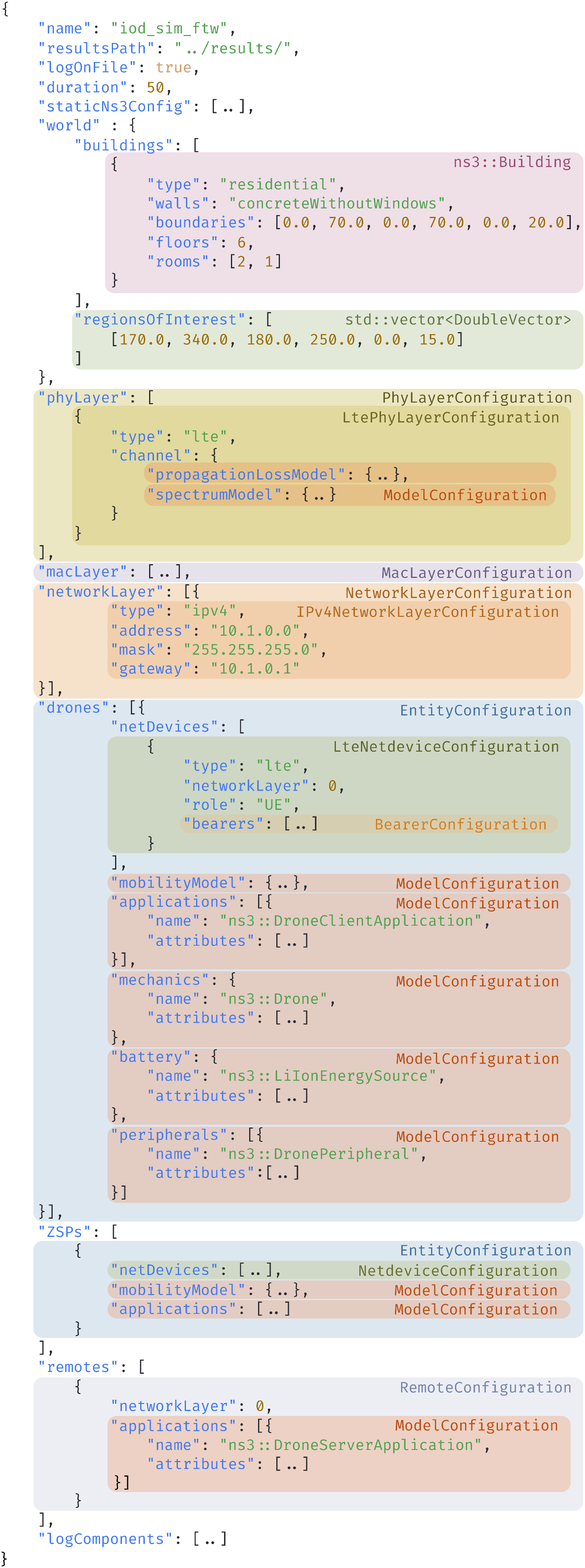}
    \caption{An excerpt of scenario configuration with an overlay of the models associated to the analyzed parts.}
    \label{fig:parsedSchema}
\end{figure}

In this work, the following configuration schema has been chosen for the General Purpose Scenario:
\begin{itemize}
    \item \texttt{name} -- A mandatory string representing the scenario name.
    \item \texttt{dryRun} -- An optional boolean to run only the semantic analyser and check that the configuration file and model setup is valid. By default, it is set to \texttt{false}.
    \item \texttt{resultsPath}-- A mandatory string representing an existing path to store simulation output files.
    \item \texttt{logOnFile} -- A mandatory boolean to output scenario logging information on a file or on standard output.
    \item \texttt{duration} -- A mandatory integer that specifies the simulation duration in seconds.
    \item \texttt{staticNs3Config} -- A mandatory array of objects, each with \texttt{name} and \texttt{value} strings, to address \ac{ns-3} static configuration parameters. The array can be empty.
    \item \texttt{world} -- An optional object containing the description of the simulated space, in particular whether to place buildings or regions of interest.
    \item \texttt{phyLayer} -- A mandatory array of objects, each representing a PHY layer configuration to be used in the scenario. Each PHY object declares its \texttt{type}, which is a mandatory string. The chosen type must be supported by the semantic analyser. Additional parameters are specific to the kind of PHY layer being configured, most notable are the chosen propagation delay model and the propagation loss model.
    \item \texttt{macLayer} -- Its description is similar to \texttt{phyLayer}.
    \item \texttt{networkLayer} -- Its description is similar to \texttt{phyLayer}.
    \item \texttt{drones} -- A mandatory array of objects, each representing a drone to be simulated. A drone requires the following properties to be configured: at least one \texttt{netDevices} in order to link it to a protocol stack and setup its network address assignment, a \texttt{mobilityModel} according to the ones available on \ac{IoD-Sim}, at least one \texttt{application} that can be installed on a drone, a \texttt{mechanics} to define mechanical properties, and a \texttt{battery}. Optionally, a \texttt{peripherals} array can also be specified in order to equip I/O devices to the drone with a specific \texttt{PowerConsumption} indication. They may also be activated by specifying the region of interest through \texttt{RoITrigger} parameter.
    \item \texttt{ZSPs} -- Its description is similar to \texttt{drones}.
    \item \texttt{remotes} -- A mandatory array of objects, each representing a remote that is described by its set of \texttt{applications}.
    \item \texttt{logComponents} -- A mandatory array of strings to enable log components available in \ac{IoD-Sim}.
\end{itemize}
An example of \ac{JSON} configuration file that realizes a simple scenario is shown in Figure \ref{fig:parsedSchema}.

\section{Simulation Development Platform}\label{sec:interfaces}
The \textit{Scenario Configuration Interface}, discussed in the previous Section, eases the design and configuration of complex scenarios from a high-level perspective. Indeed, \ac{JSON} greatly facilitates management and maintainability thanks to its dry and human-readable syntax. However, user experience is still hindered for the following reasons:
\begin{itemize}
    \item As \ac{IoD-Sim} grows in size and introduces complex and powerful models over time, the learning curve to effectively use this simulator steepens, which leads to a general \ac{QoE} degradation.
    \item This project is continuously developed and upgraded with new features, technologies, standards, and more advanced behaviors. A high-level abstraction helps reducing the barrier for scenario developers in approaching new features and the required effort, time, and overhead to implement a scenario.
    \item A general purpose configuration interface, provided in the form of \ac{JSON}-encoded files, does not give any visual clue on scenario design. Indeed, plain text files alone require low-level knowledge of the simulator, thus implying that the user has to rely on its experience and imagination to effectively know all the aspects related to a complex scenario configuration, such as the number of drones, their trajectories, their purpose, their equipment, the topology of the ground infrastructure, and the services exposed by remote nodes.
    \item Error reporting messages cannot be easily understood by end-users, forcing the use of a debugger to isolate the problem. Therefore, a semantic analysis would be beneficial to detect problems at scenario configuration.
\end{itemize}
To address all the points above, \ac{IoD-Sim} \textit{Simulation Development Platform} provides a set of extensions, i.e., the \textit{Report Module}, output files for data analysis, and standalone applications for scenarios design, such as \textit{Airflow}. These tools ease scenario design and analysis, thus ensuring that \ac{IoD-Sim} can be easily introduced to newcomers, especially university students and researchers.

\subsection{Report Module}\label{sec:report_module}
The \textit{Report Module} is an extension of \ac{IoD-Sim} which stores data at run-time and elaborates, at the end of simulation, a comprehensive summary.
The aim of the extension is to introspect simulator's data structures to gather relevant data to be reported (e.g., data traffic, trajectory, and telemetry).
To provide a final report that is both human and machine readable, the \ac{XML} format has been chosen. Therefore, a schema is defined to describe the expected structure of the produced file.

More insights about the structure of the proposed extension are provided hereby.
The root \ac{XML} element, i.e, \verb|Simulation|, represents the summary of a scenario previously executed.
The attributes that characterize the simulation are \texttt{scenario}, which is a string that carries the name of the scenario that was executed, and \texttt{executedAt}, which reports the date and time of execution of this simulation.
Moreover, \verb|Simulation| presents further information about simulation results, such as its \texttt{duration}, which is reported in \texttt{real} and \texttt{virtual} time, \texttt{World}, which contains the \texttt{Buildings} and \texttt{InterestRegions}, and entities containers.

The first of these containers is \verb|Zsps|, which is a complex \ac{XML} type that summarizes each \ac{ZSP} through \verb|position| described by the 3D coordinates, and \verb|NetDevices|, which is a list of configured network devices. Each of them is described by structures that represent the configuration of the PHY, MAC, and network layers, together with the data traffic. Each captured packet is expressed by direction, length in bytes, timestamp, and textual representation of the payload.

Similarly, \verb|Drones| summarizes the state of each \verb|Drone|. This structure maintains the \verb|NetDevices| already discussed for \verb|Zsps|. Additionally, particular characteristics of drones are reported, such as \verb|trajectory| and the set of onboard \verb|Peripherals|. The former is defined by a list of points, each of them with its own timestamp. The latter reports the characteristics of the used peripherals type.

Finally, \verb|Remotes| are described only by their \verb|NetDevices|.

This output \ac{XML} file is put together with other files relevant to the simulation in the \verb|results| directory.

\begin{figure}[!ht] 
    \centering
    \includegraphics[width=0.8\columnwidth]{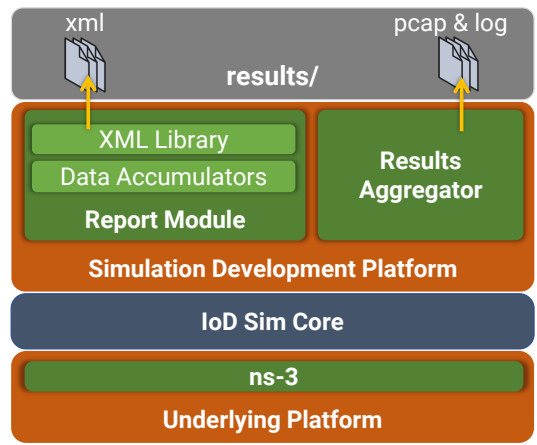}
    \caption{Block diagram of the Report Module.}
    \label{fig:ReportModule}
\end{figure}

\subsection{Results Aggregator}
\textit{Log Files} gather all the relevant information and debug messages of simulation internal components. Primarily, the \textit{General Purpose Scenario} emits \texttt{progress.log} and \texttt{IoD Sim.log} files. The former is the output of the progress information messages that are also delivered on the standard output during scenario execution. The latter contains all debug messages coming from different internal components of \ac{IoD-Sim}. The log components can be enabled by specifying them in the \texttt{logComponents} field of the scenario configuration \ac{JSON} file.

\texttt{progress.log} file starts by determining the current date and time of the start of the simulation. For each second, it prints a status report on a single line.
The status report presents the following fields:
\begin{itemize}
    \item The simulation time instant at which the report is referring to.
    \item The speed up in simulating the scenario in respect to the real time. This is dependent on the simulator performances and how many events are elaborated.
    \item The number of events processed in the time interval relative to the previous status report.
\end{itemize}
The file then ends with the current date and time and the duration of the simulation as \textit{Elapsed wall clock}.

\textit{Trace Files} are ASCII-encoded text files that record all the activities regarding a specific Network Device. All the traces are bounded on what actually is sent or received at the MAC layer. A \textit{Trace File} name is composed by three fields, separated by hyphen: (i) the global layer name, (ii) the unique identifier of the host in the network, and (iii) the unique identifier of the host network device.
For instance, \texttt{internet-2-1.tr} indicates that the trace has been done on the first network device of the second host in the virtual Internet network.

\textit{\ac{LTE} Log Files} are ASCII-encoded text files that represent a series of statistics on relevant \acp{KPI} of \ac{LTE}. These log files are focused on specific low-level layers of the \ac{LTE} stack, particularly PHY, MAC, Radio Link Control, and Packet Data Convergence Protocol.
For each layer, there are two separate trace files: one for \textit{downlink} and one for \textit{uplink} communications.
As part of the \textit{\ac{LTE} Log Files}, there are also PCAP traces of the \textit{S1-U} interface that links the \ac{RAN} with the Evolved Packet Core.

\textit{PCAP Files} are well-known files that record network activity in the PCAP format and contains the traffic occurred on a certain network device of a host. The filename format is the same of \textit{Trace Files}.
Due to the fact that these files are binary, a suitable decoder should be used to explore the data structure.
A popular decoder is \texttt{libpcap} open source project, used by frameworks for PCAP data analysis, e.g., \textit{Scapy}, and \ac{GUI} programs such as \textit{Wireshark}.
As these \textit{PCAP Files} are generated by a simulation, each captured frame is marked with the relative timestamp of the simulation. Therefore, each \textit{PCAP File} starts with the transmission/reception of captured frames at $0$ seconds.

\begin{figure}[!t]
    \centering
    \includegraphics[width=\columnwidth]{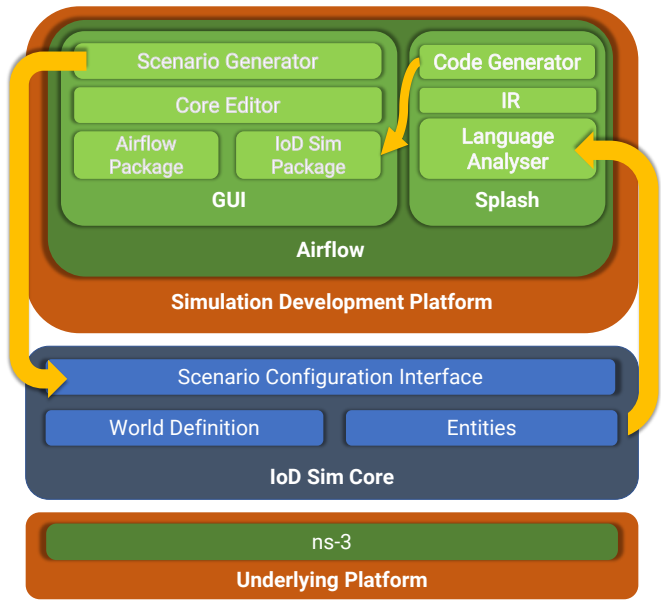}
    \caption{Airflow Architectural Design.}
    \label{fig:airflow-architecture}
\end{figure}
\begin{figure}[!t]
    \centering
    \includegraphics[width=\columnwidth]{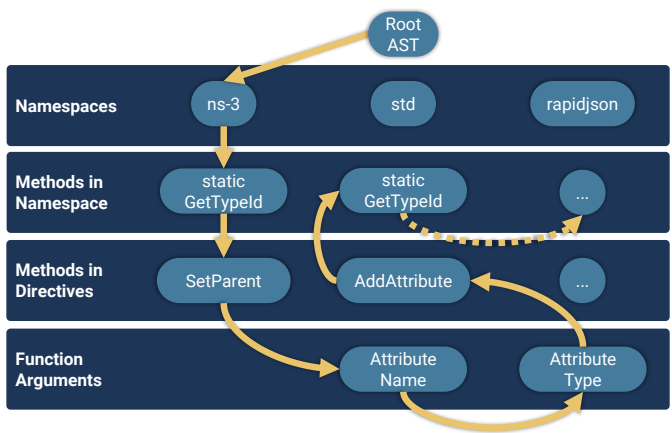}
    \caption{The tree-traversal search algorithm employed by Splash to extract the models from the \acs{IoD-Sim} source code. The numerical ordering given on the edges reflects the algorithm logic used to extract model information.}
    \label{fig:ast-traversal-algo}
\end{figure}
\begin{figure*}[!htb]
    \centering
    \includegraphics[width=\textwidth]{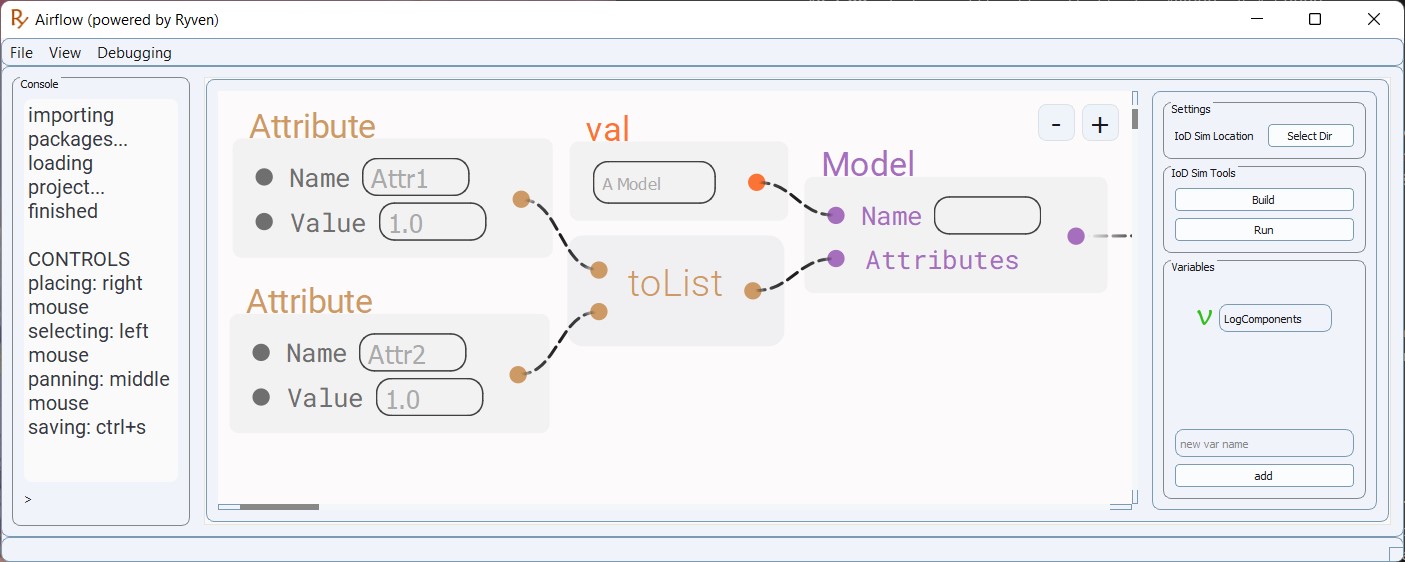}
    \caption{An overview of the configuration of a generic model in Airflow.}
    \label{fig:airflow-block-example}
\end{figure*}

\subsection{Airflow}\label{sec:scenario_development}
Airflow is a high-level abstraction tool that gives visual clues during simulation design, thus enriching the user experience especially for newcomers.
Airflow has been developed on top of Splash, a specialized transpiler for \ac{IoD-Sim}. It scans the source code of the simulator and outputs visual blocks that can be referenced in the \textit{Core Editor} to configure a scenario. Thanks to the \ac{GUI} editor, a scenario can be exported into a \ac{JSON} file that can be interpreted by \ac{IoD-Sim} \textit{Scenario Configuration Interface}.
From a software design standpoint, as illustrated in Figure \ref{fig:airflow-architecture}, the Airflow project is entirely decoupled from \ac{IoD-Sim}. Its integration with the simulator relies on interfaces that enable bidirectional communications.

\subsubsection{Splash}
Splash is a middleware that analyzes \ac{IoD-Sim} source code and translates \ac{ns-3} models into visual blocks code used by Airflow.
These blocks can be added to the editor as external packages. Splash enables the decoupling mechanism able to ensure that Airflow and \ac{IoD-Sim} can be developed asynchronously and updated when needed.

In particular, it accomplishes the following tasks:
\begin{enumerate}
    \item Parses the source code of \ac{IoD-Sim} by relying on Clang lexical and syntax analyzers, producing the \ac{AST} that is stored into a binary file.
    \item Scans the \ac{AST} to find relevant simulation models, excluding internal structures and routines that are not relevant for the design of a scenario. This information is then encoded in an \ac{IR}.
    \item Optimizes the \ac{IR} by solving model hierarchies and by removing redundancies.
    \item Generates Python code that describes the models as Airflow visual blocks. This output can then be moved to the Airflow project folder for integration.
\end{enumerate}
Concretely, this pipeline works as follows. The script \texttt{splash.sh} can be executed by passing the \ac{IoD-Sim} project directory as argument. The program then searches for any relevant C++ source code files in it. This process is eased by the \ac{ns-3} convention: models have the suffix \texttt{-model.cc}, \texttt{-manager.cc}, \texttt{-mac.cc}, and \texttt{-application.cc} in their filenames. To this end, other files are filtered out to optimize parsing operations and preventing the exposure of simulator's internal structures.

For each file found, the \texttt{clang} command is used to analyze the source code and solve any include directives needed by the preprocessor. Finally, the output is an \ac{AST}, which is encoded in an optimized binary file readable only through \texttt{clang}'s \acp{API}. The file extension is named \ac{PCH}.

The \ac{PCH} file is then passed into \texttt{splash} core executable. This program relies on \texttt{cxxopts} library to behave like an interactive command-line application, on \texttt{boost-json} to serialize C++ data structures in \ac{JSON}, and on \texttt{libclang} to read the \ac{AST}.
The application requires the \ac{PCH} file path as input with the output directory path in order to store the \acp{IR}. These \ac{IR} files are encoded into \ac{JSON} to ensure software interoperability and readability.

Once the command-line program is executed, the entire translation unit of the \ac{AST} is scanned in order to lookup for any model used in the simulator.
A custom tree-traversal algorithm is used to optimize the parse time.
It works as an hybrid implementation of the classical \textit{Breadth-first} and \textit{Depth-first} search algorithms. A high-level representation of the translation unit is given in Figure \ref{fig:ast-traversal-algo}. The key feature of this approach is the speed up introduced by the algorithm. In fact, it first traverses the tree using \textit{Depth-first} to find the depth at which one or more \texttt{ns3::TypeId} can be found and then uses \textit{Breadth-first} to analyze each model at the same depth.
The same strategy is applied to extract all the attributes relevant to the simulator model. Each model is represented and exported into a \ac{JSON} file having the following structure: the name of the parent model, the model name, and a list of attributes, each one described by a name, an optional description, and the \ac{ns-3} data type that characterizes it.
Once the entire model hierarchy is solved and optimized, the attributes are copied from parent to children. Then, a code generator is executed to create the visual blocks for the editor \ac{GUI}. Each block name reflects the model's one and the attributes are considered as block input parameters. The generated Python code is interpreted by the \ac{GUI} to display a visual block with the model name as its title and model attributes as its inputs, as illustrated in Figure \ref{fig:airflow-block-example}.

\begin{figure*}[!htb]
    \centering
    \includegraphics[width=\textwidth]{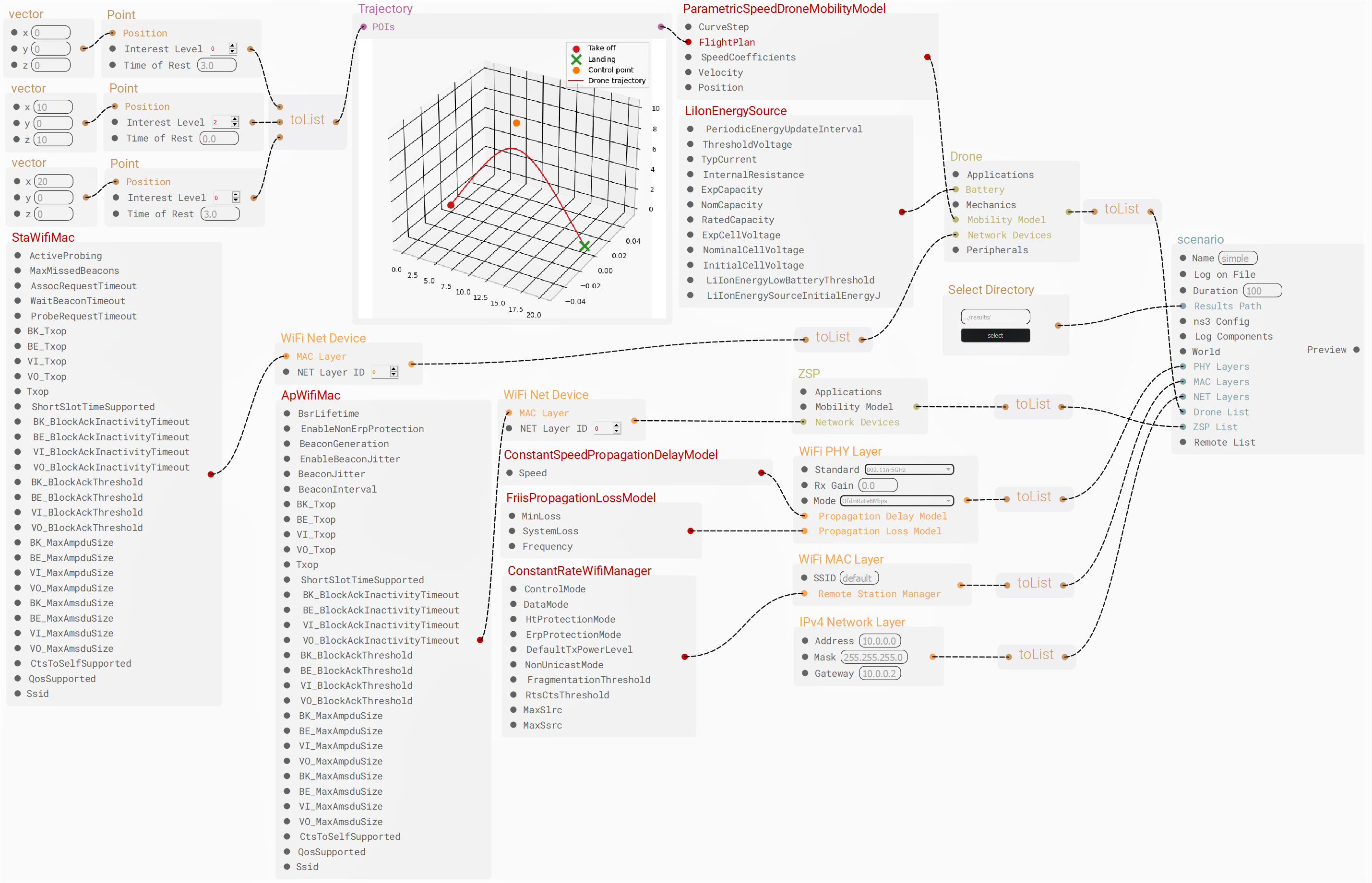}
    \caption{A simple scenario with one drone and \ac{ZSP} designed from scratch in Airflow.}
    \label{fig:airflow-scenario-simple}
\end{figure*}

\subsubsection{\acl{GUI}}
The Airflow \ac{GUI}, shown in Figure \ref{fig:airflow-block-example}, is based on the open source Ryven\footnote{\url{https://ryven.org/}} engine, which is a dynamic runtime, \textit{flow-based} visual programming environment for Python scripts.
It offers: (i) a central rendering view to place blocks and link them together, (ii) a settings area to customize options, (iii) a variable management section to include and store data that can be integrated with the flow, and (iv) a console to report errors. Ryven includes additional features to optionally debug internal routines with the help of console messages.
Moreover, thanks to its modular design, it allows blocks generated by Splash to be aggregated into \textit{packages}.
Ryven has been deeply extended to inter-operate with \ac{IoD-Sim}, especially for its compatibility with the \textit{Scenario Configuration Interface}.

The user interface is organized into the following components:
\begin{enumerate}
    \item A menu bar at the top of the \ac{GUI} window.
    \item A \textit{Console} on the left in order to monitor errors and messages coming from Airflow or \ac{IoD-Sim}. Informative messages are reported in blue, while errors are displayed in red.
    \item A central workspace to design the scenario by placing blocks and connecting them together.
    \item A settings panel on the right.
\end{enumerate}

The menu bar is divided into three categories: with \texttt{File} it is possible to import Airflow packages to extend the user experience with third-party visual blocks. Moreover, it provides features to save the project or export it as \ac{IoD-Sim} configuration file.
\texttt{View} offers aesthetic options, such as changing the theme, make a screenshot of the project, and tune performance parameters.
Finally, \texttt{Debugging} enables technical features to ease troubleshooting of the program, such as increasing verbosity level on the \textit{Console}.

The central workspace is the canvas where blocks and links are placed by the user to design a scenario. A block, as depicted in Figure \ref{fig:airflow-block-example}, consists of a set of inputs and outputs. Each input and output can be connected to other outputs and inputs of other blocks, in order to create a tree. The root block is named \textit{Scenario}.
Each block has a different meaning and function. As a general overview, blocks can be divided into the following categories: operators, helpers, and \ac{IoD-Sim} models. Operators are built-in blocks that can be used to work with values, constants, and data structures. Instead, helpers are special blocks that eases the configuration of a scenario, i.e., entities, Wi-Fi, and \ac{LTE} configuration blocks.
Usually, blocks provide a single output without label. This output delivers the information of the block, along with all its inputs, to the next connected block.
Blocks can be added to the workspace by a specific menu that is shown by clicking with the right mouse button.
Moreover, each block can be right-clicked to show its contextual menu that can be used (i) to remove it, (ii) to refresh it (and hence reading again all its inputs), and (iii) to use some particular features available in certain blocks. For instance, \texttt{toList} offers some additional controls to add or remove inputs.

In the settings panel, it is possible to set the \ac{IoD-Sim} path in order to enable interoperability features, such as to check the scenario configuration for errors, or to run the scenario and report the status on the \textit{Console}. These features can be used by clicking on the \texttt{Build} and \texttt{Run} buttons, respectively.
Finally, a variable manager can be used to create, store, and later reference values by their respective labels on the workspace. This allows to reduce redundancy and to make the block tree more compact.

\section{Simulation Campaign}\label{sec:simulations}
This Section demonstrates the huge potential of \ac{IoD-Sim} by means of an extensive simulation campaign which investigates the many facets of \ac{IoD} scenarios.
Firstly, the discussion explains how the simulation can be designed.
Secondly, three different scenarios with increasing complexity are presented.

In particular, the first scenario discusses the use-case of telemetry with few drones flying in a \ac{RoI}, which follow customized trajectories while gathering data.
The purpose of this scenario is to demonstrate that it is possible to monitor one or more variables with on-board sensors, while estimating the energy consumption associated with flight dynamics.

The second scenario has a wider perspective, since it focuses on surveying and monitoring activities, further completed with the acquisition of multimedia signals by each drone.
The possible applications include several real-world use cases in the fields of civil engineering, smart agriculture, or in environmental monitoring, e.g., coastal erosion and other slow phenomena.
In fact, in this scenario, drones are on a mission in neighboring areas, since it is assumed that the information of interest need to be contextualized, i.e., must be gathered at the same time. Furthermore, this case investigates the possibilities enabled by different data storage capabilities of by drones. Also, the offloading functionality, of the acquired data, avoids the overload/saturation of onboard available resources. Once data is gathered, they can be involved in offline post-processing, evaluation, and analysis.

The third scenario has been specifically designed to be the reference benchmark for \ac{IoD} applications. It is settled in the context of smart cities, and it involves clusters of low-power \ac{IoT} sensors. This scenario models real-world applications and, hence, shadowing and pathloss phenomena are included, thanks to the adoption of propagation models that are influenced by the presence of buildings. In order to guarantee a reliable communication, drones are in charge of relaying traffic to ensure coverage to all sensors in the city.

\subsection{Scenario Design}
Airflow represents the foremost application for visual scenario development.
To better understand how to design simulations, a simple configuration setup is provided hereby.
The envisioned scenario considers a drone that follows an arc-like trajectory and communicates telemetry to a \ac{ZSP} by means of Wi-Fi. Specifically, the drone acts as a \textit{station} and the \ac{ZSP} as an \textit{access point}.
The entire configuration is depicted in Figure \ref{fig:airflow-scenario-simple}, where all the visual components, encompassed in the Airflow workspace, are properly set up and linked together. Starting from the right, the block \texttt{Scenario} glues some configuration input values, e.g. \texttt{Name} and \texttt{Duration}, with more complex components, such as (i) \texttt{PHY/MAC/NET Layers}, (ii) \texttt{Drone List}, and (iii) \texttt{ZSP List}.

In particular, the communication layers are configured to implement the Wi-Fi stack.
The \texttt{WiFi PHY Layer} object defines the PHY layer to be used with particular propagation and loss models. The \texttt{WiFi MAC Layer}, instead, specifies the \ac{SSID} of the network and the Wi-Fi Manager object that handles MAC control plane. Further, the \texttt{IPv4 Network Layer} determines the address and mask of the overlying network.

Both \texttt{Drone List} and \texttt{ZSP List} properties are connected to the simulated entities, namely \texttt{Drone} and \texttt{ZSP}. These two components share different properties such as \texttt{Applications}, \texttt{Mobility Model} and \texttt{Network Devices}. However, the \texttt{Drone} block is characterized also by its unique features, i.e., \texttt{Peripherals}, \texttt{Mechanics}, and \texttt{Battery}.
In this configuration, the \texttt{ConstantPositionMobilityModel} allows to place the \texttt{ZSP} to a fixed location, while the \texttt{ParametricSpeedMobilityModel} is employed to define the drone trajectory.
In this regard, the \texttt{Trajectory} component, linked to the \texttt{FlightPlan} property of the mobility model, facilitates the design of the desired path.

Both drone and \ac{ZSP}'s \texttt{Network Devices} property is linked to a \texttt{WiFi Net Device} block. While \texttt{StaWifiMac} characterizes the device of the former, \texttt{ApWifiMac} is associated to the latter.
Finally, a \texttt{LiIonEnergySource} defines the power supply of the drone.

The development strategy discussed above represents the common ground for the design of the following three scenarios.

\begin{figure}[!ht]
    \centering
    \includegraphics[width=\columnwidth]{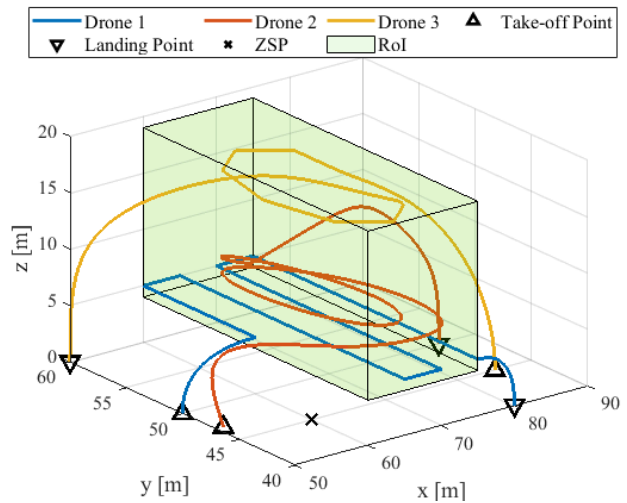}
    \caption{Scenario \#1.}
    \label{fig:s1-trajectories+roi}
\end{figure}
\begin{figure}[!ht]
    \centering
    \includegraphics[width=\columnwidth]{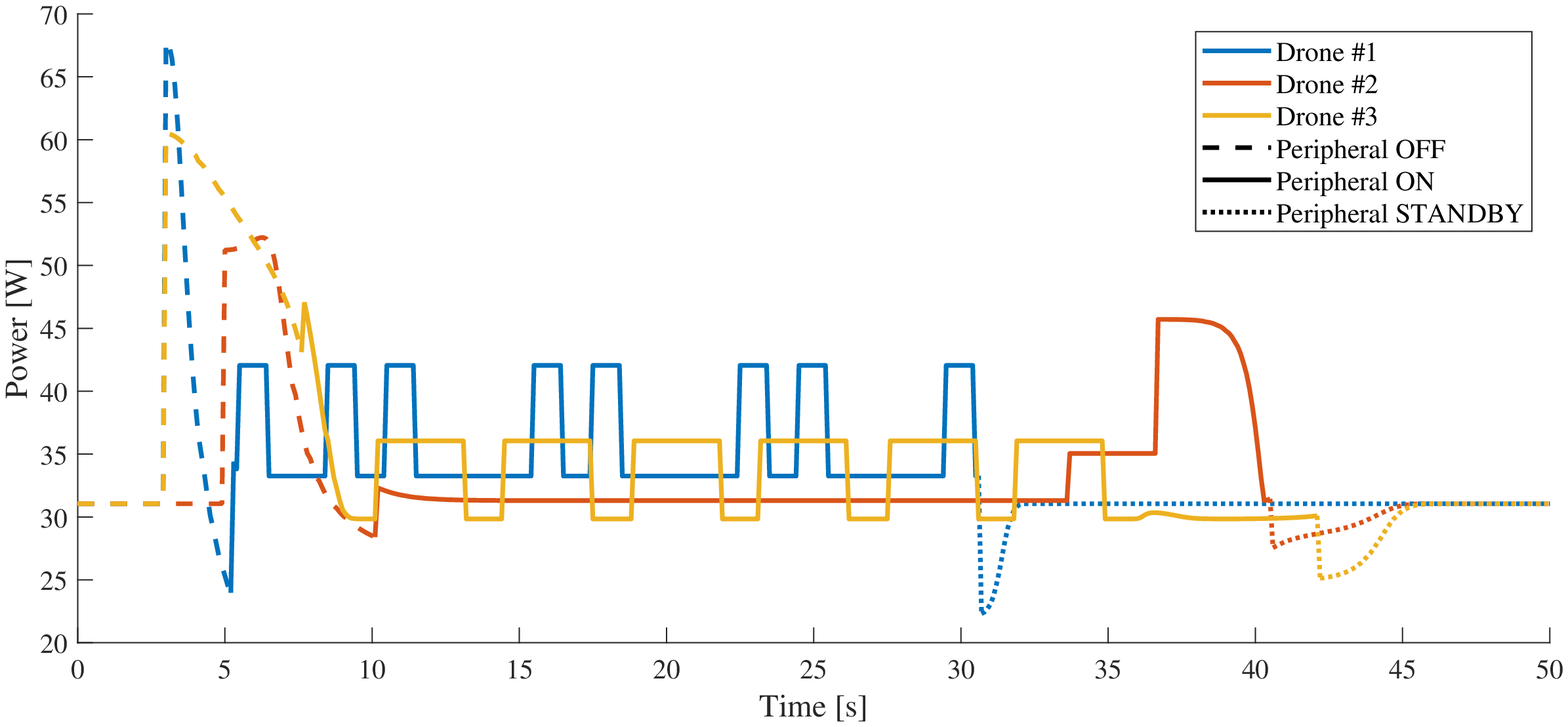}
    \caption{Power consumption and peripheral state for each drone, in the first scenario.}
    \label{fig:s1-total-power-consumption-drones}
\end{figure}
\begin{figure*}[!ht]
    \centering
    \begin{subfigure}[h]{0.33\textwidth}\hfill
        \centering
        \includegraphics[width=\columnwidth]{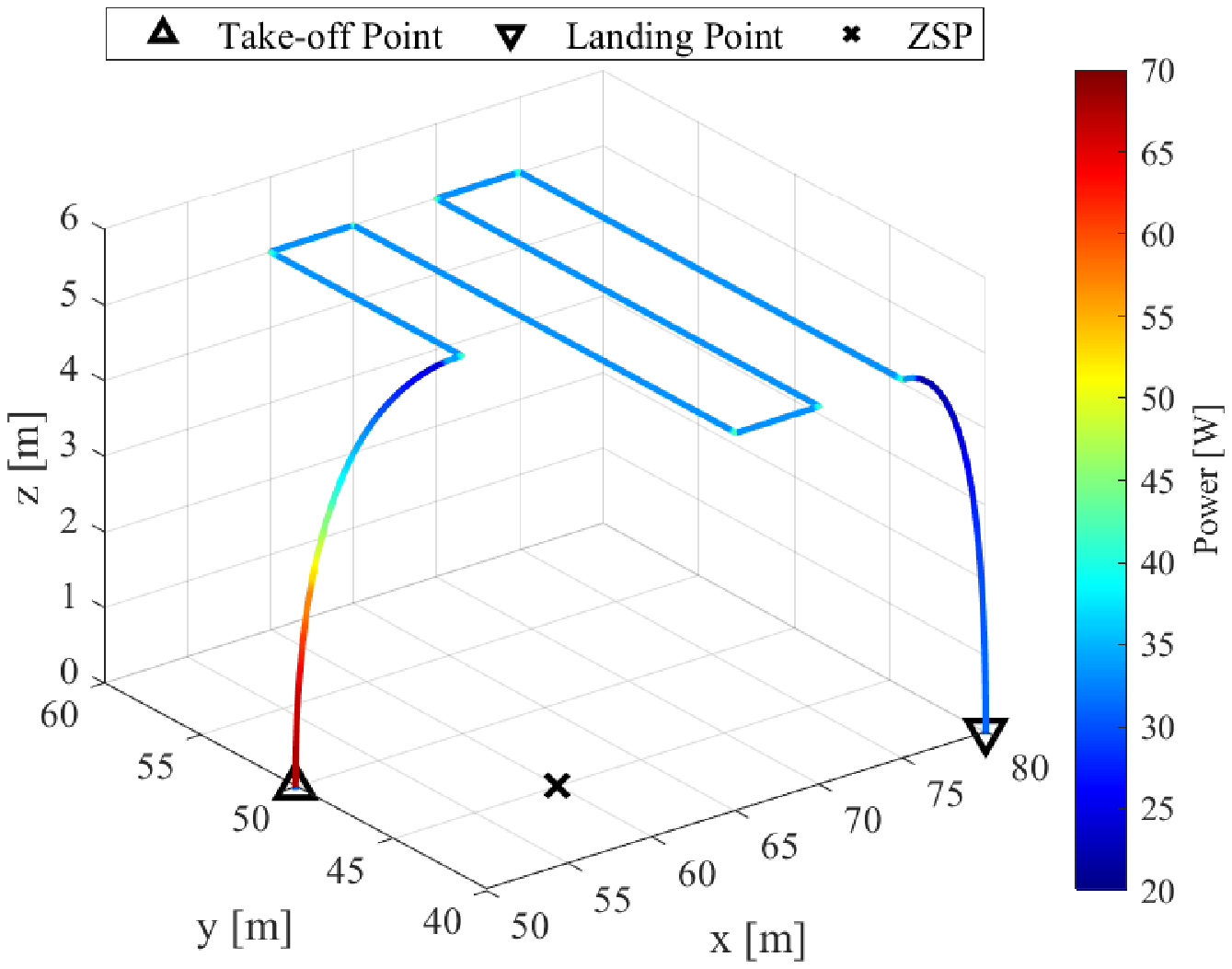}
        \caption{Drone \#1.}\label{fig:drone-traj-1st}
    \end{subfigure}%
    \begin{subfigure}[h]{0.33\textwidth}
        \centering
        \includegraphics[width=\columnwidth]{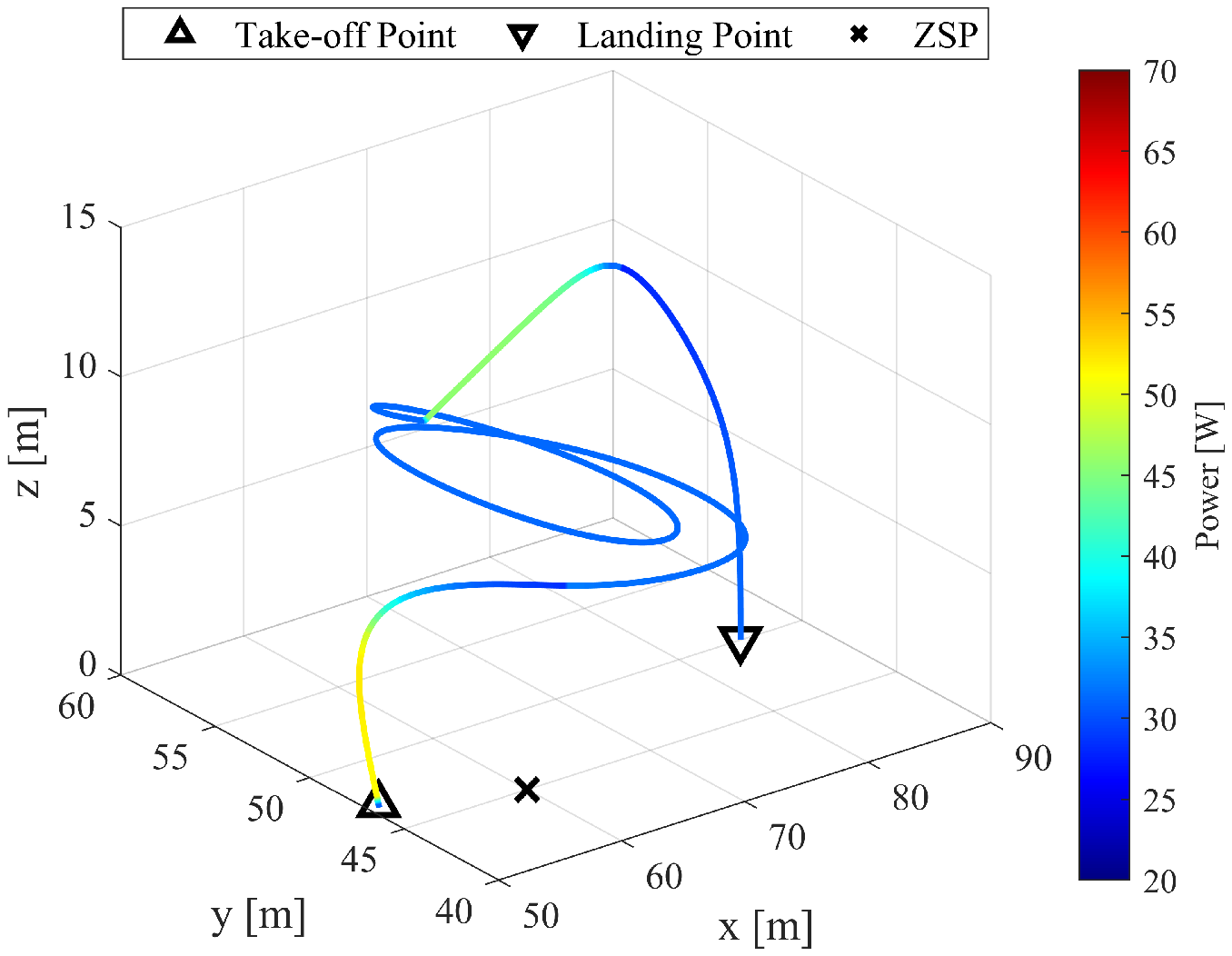}
        \caption{Drone \#2.}\label{fig:drone-traj-2nd}
    \end{subfigure}%
    \begin{subfigure}[h]{0.33\textwidth}
        \centering
        \includegraphics[width=\columnwidth]{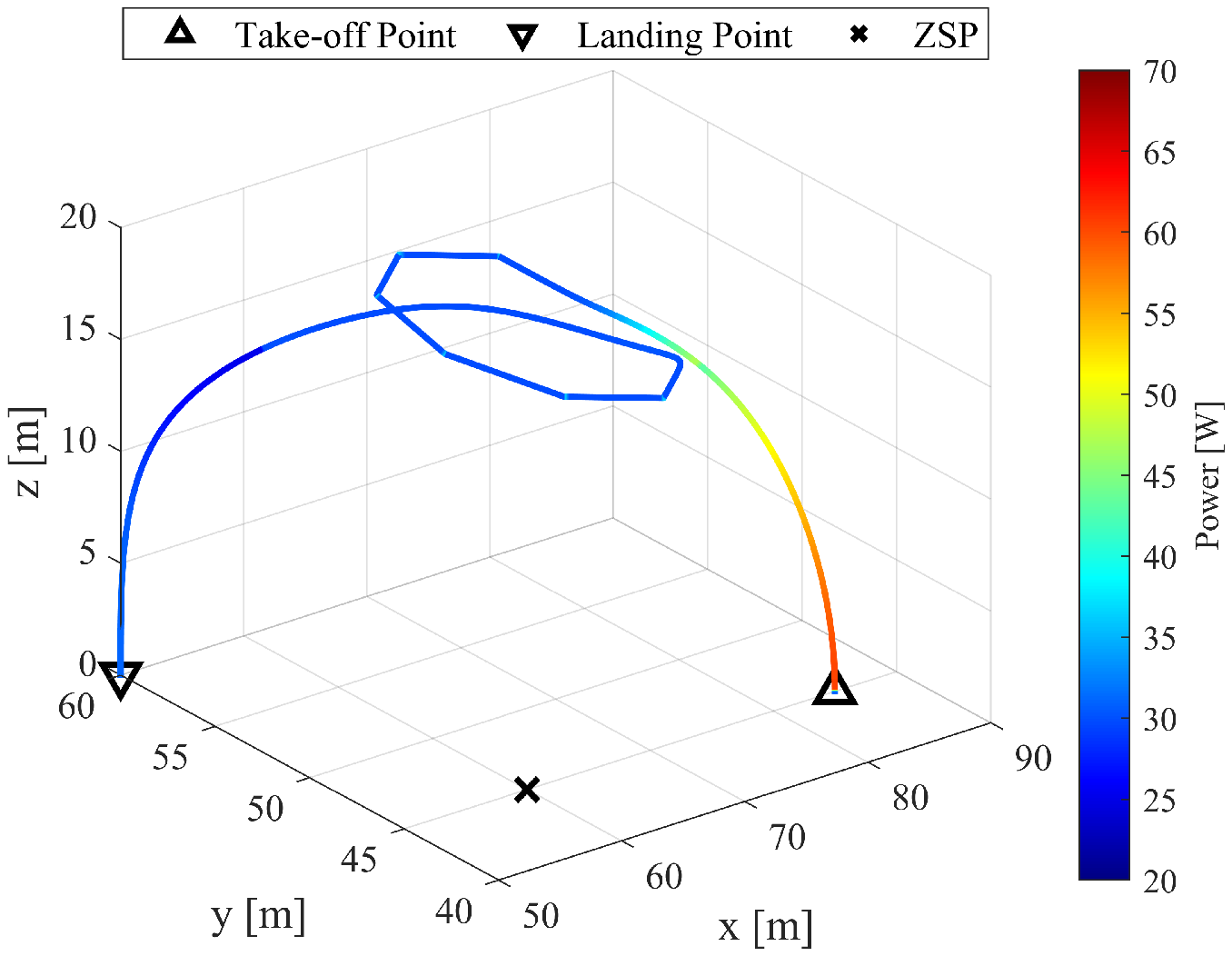}
        \caption{Drone \#3.}\label{fig:drone-traj-3rd}
    \end{subfigure}
    \caption{Drones' trajectories with their power consumption, in the first scenario.}
    \label{fig:drone-traj-1st-scenario}
\end{figure*}
\begin{figure}[!ht]
    \centering
    \includegraphics[width=\columnwidth]{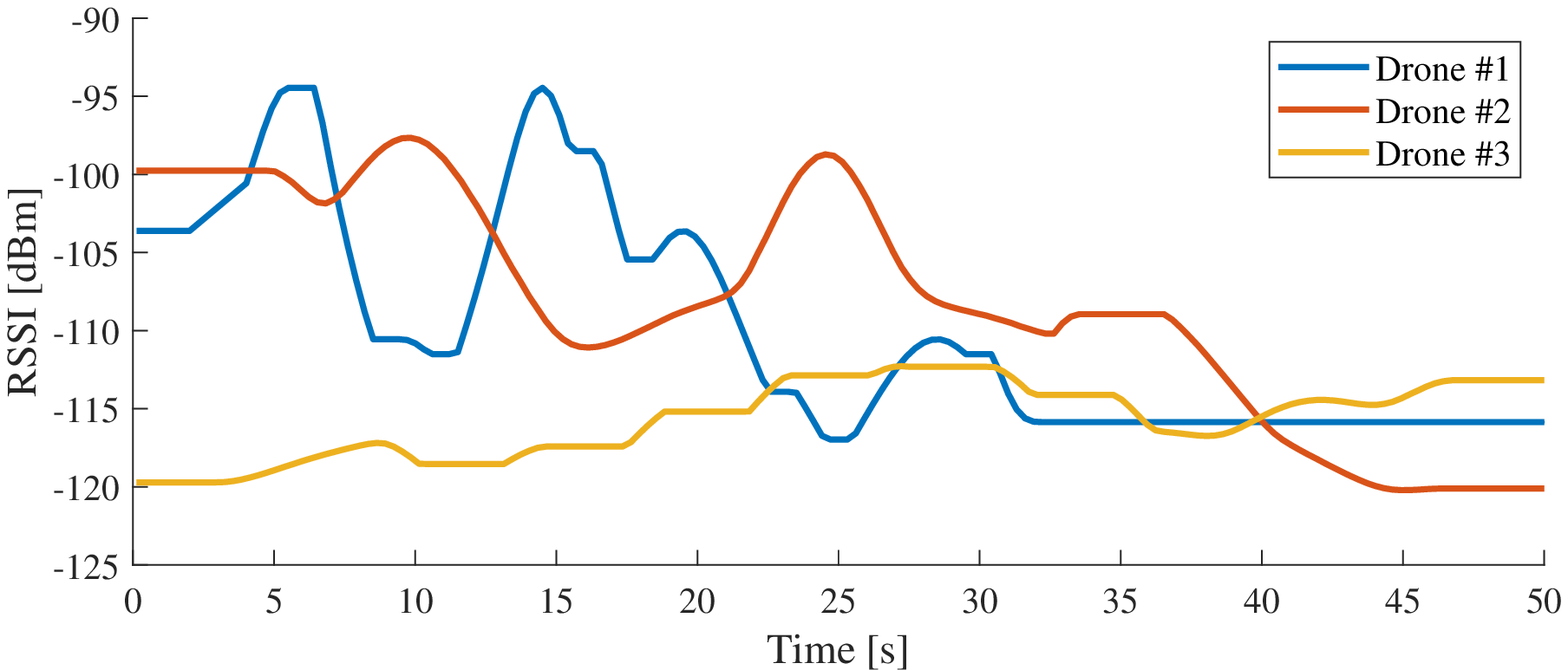}
    \caption{\ac{RSSI} of each drone in the first scenario.}
    \label{fig:s1-rssi-drones}
\end{figure}

\subsection{Scenario \#1 - Telemetry}

The first scenario, as depicted in Figure \ref{fig:s1-trajectories+roi}, envisions three drones with the same mechanical characteristics, all equipped with an \ac{IMU}. In this scenario, drones are flying in the same \ac{RoI}, at a constant speed, following different trajectories. Moreover, a \ac{ZSP} is deployed on the ground. The latter is released in $[60 \ 45]^T$, which continuously monitors drones' operations by acquiring telemetry through Wi-Fi.

\acp{UAV}' trajectories are based on the ParametricSpeedDroneMobilityModel, which is configured to guarantee a constant speed of $\SI{5}{m/s}$, $\SI{3}{m/s}$, and $\SI{4}{m/s}$, respectively.
They are also equipped with \acp{IMU} which are generic drone peripherals that provide basic telemetry data to the \ac{ZSP} thanks to a dedicated application, as mentioned in Section \ref{sec:telemetry_applications}. It is worth specifying that drones' \acp{IMU} have different power consumption, i.e., $\SI{12}{W}$, $\SI{5}{W}$, and $\SI{6}{W}$.

The outcome of the simulation is hereby discussed.
Figures \ref{fig:s1-total-power-consumption-drones} and \ref{fig:drone-traj-1st-scenario} depict the power consumption trend with respect to time and trajectories.
In the former, the three curves share an initial peak which corresponds to the energy required to take-off. Indeed, acquiring altitude requires more power than flying along the xy plane, as highlighted. This phenomenon is further remarked in Drone \#2 landing maneuver. It includes a little parabola that yields a peak in the last part of the associated curve of Figure \ref{fig:s1-total-power-consumption-drones}, which is also present in Figure \ref{fig:drone-traj-2nd}.
After $\sim\SI{10}{s}$, the drones reach and almost maintain a target altitude. The corresponding power consumption, for Drones \#1 and \#3, is characterized by peaks due to hovering over the interest points for  $\SI{1}{s}$ and $\SI{3}{s}$, respectively. These points are identified by the vertices of the snake-like and octagon-shaped trajectories.
Instead, this phenomenon is not present on Drone \#2, since its trajectory describes a continuous curve.
\begin{figure}[!ht]
    \centering
    \includegraphics[width=\columnwidth]{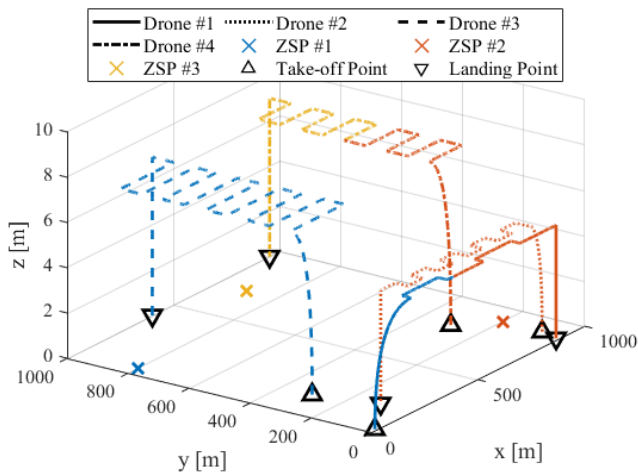}
    \caption{Trajectory design and eNB attachment for each drone, in the second scenario.}
    \label{fig:scenario-2}
\end{figure}
When the drones enter the \ac{RoI}, the peripherals become active and hence the \acp{IMU}' power contribution is non-zero. It can be noticed as spikes in the curves of Figure \ref{fig:s1-total-power-consumption-drones}, especially in Drones \#1 and \#2, since they are equipped with two more energy-demanding peripherals. As soon as drones exit such region, the peripherals go into standby mode, which preserves energy.

Figure \ref{fig:s1-rssi-drones} illustrates the measured \ac{RSSI} of each drone at the \ac{ZSP}. It clearly emerges that, on average, Drones \#1 and \#2 maintain a better signal quality with respect to the \ac{UAV} \#3. Obviously, the higher altitude, and hence the greater distance from the \ac{ZSP}, worsens the communication quality, due to the Friis propagation loss employed to model the fading effects in this scenario.

\begin{figure}[!ht]
    \centering
    \includegraphics[width=\columnwidth]{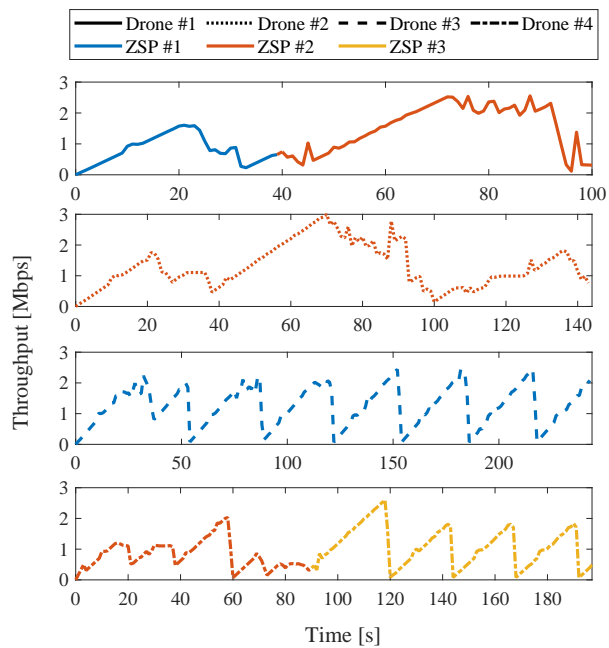}
    \caption{Drones' throughput, in the second scenario.}
    \label{fig:scenario-2-lte-throughput}
\end{figure}

\subsection{Scenario \#2 - Multimedia Signals Acquisition}
The second scenario, as depicted in Figure \ref{fig:scenario-2}, involves a swarm composed by four drones in charge to acquire multimedia signals at different data rates which are then stored on-board and off-loaded to a remote server. To allow data upload, three \acp{ZSP}, also referred to as eNB, are deployed on the ground at $[50 \ 800]^T$, $[900 \ 200]^T$, and $[700 \ 900]^T$, respectively.
\begin{figure}[!ht]
    \centering
    \includegraphics[width=\columnwidth]{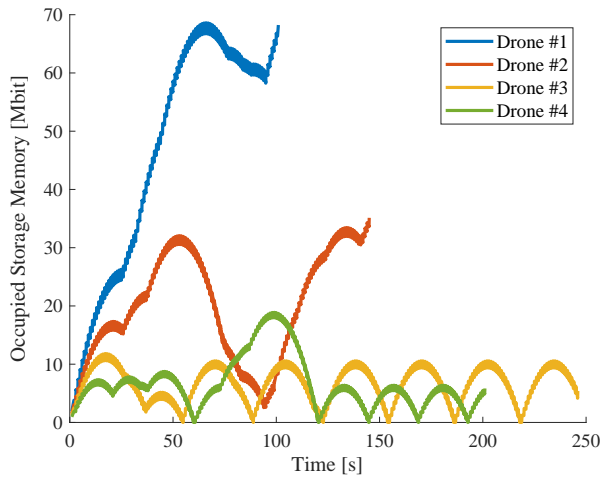}
    \caption{Memory occupancy for each drone, in the second scenario.}
    \label{fig:scenario-2-storage}
\end{figure}
All the entities involved in the mission, which lasts $\SI{250}{s}$, are equipped with \ac{LTE} interfaces, where the Okumura-Hata propagation loss model has been employed. Drones follow snake-like trajectories, each different from the other in terms of amplitude and frequency. Nevertheless, they adopt the same mobility model with a constant acceleration of $\SI{4}{m/s^2}$ and a maximum velocity between $15$ and $\SI{20}{m/s}$. Moreover, they are equipped with cameras that acquire at $\SI{2}{Mbps}$, $\SI{1.6}{Mbps}$, $\SI{1.3}{Mbps}$, and $\SI{1}{Mbps}$, respectively.
The communication between each \ac{UAV} and the remote server is handled by \textit{Generic Traffic Applications} (see Section \ref{sec:generic_traffic_applications}), with a payload size of $\SI{1024}{bytes}$ and a TCP Max Segment Size of $\SI{1380}{bytes}$.

In the same figure, it can be further observed the attachment of the drones to the \acp{ZSP}. Throughout the mission, Drones \#2 and \#3 remain linked to the same eNB, i.e., \ac{ZSP} \#2 and \#1. On the other hand, \ac{UAV} \#1 and \#4 perform a handover procedure which changes the reference \ac{ZSP} from \#1 to \#2 and from \#2 to \#3, respectively. It is worth noting that, despite Drone \#1 takes-off in the same area where Drone \#2 lands, they are not attached to the same \ac{ZSP}. Indeed, even if the two trajectories share the same direction, they have opposite verse: while one approaches a eNB, as the mission goes by, the other flies away from the \ac{ZSP} without really getting closer to another one.

Figure \ref{fig:scenario-2-lte-throughput} shows the throughput for each drone on the associated \ac{ZSP}, over time. \ac{UAV} \#1 experiences an average data rate of $\sim\SI{1}{Mbps}$, until the handover procedure takes place, which increases this value by $\sim$50\%. Similarly, the average throughput of Drone \#4 is ameliorated, since it increases from $\sim\SI{800}{kbps}$ to $\sim\SI{1.1}{Mbps}$. It is worth noting that there exists a pattern correspondence between the throughput and occupied storage curves (see Figure \ref{fig:scenario-2-storage}). This is particularly evident for Drones \#3 and \#4. When the occupied memory goes to zero, the data rate goes to zero as well. Indeed, for the information causality principle, it cannot be transmitted more information than the stored amount. Notice that this happens as long as the acquiring rate remains lower or equal to the channel capacity which, for instance, is not the case of Drone \#1.

\subsection{Scenario \#3 - Smart Cities}
The third scenario reproduces a smart city context, in which drones are in charge of relaying traffic coming from clusters of \acp{GU}, using the Wi-Fi technology, to a remote server over the Internet, through \ac{LTE}. In this regard, the presence of buildings plays an important role both in trajectory design and in fading phenomena. The envisioned scenario is designed starting from the map of an urban area in the neighborhood of the Central Station of Bari, Puglia, Italy.
\begin{figure}[!t]
    \centering
    \includegraphics[width=\columnwidth]{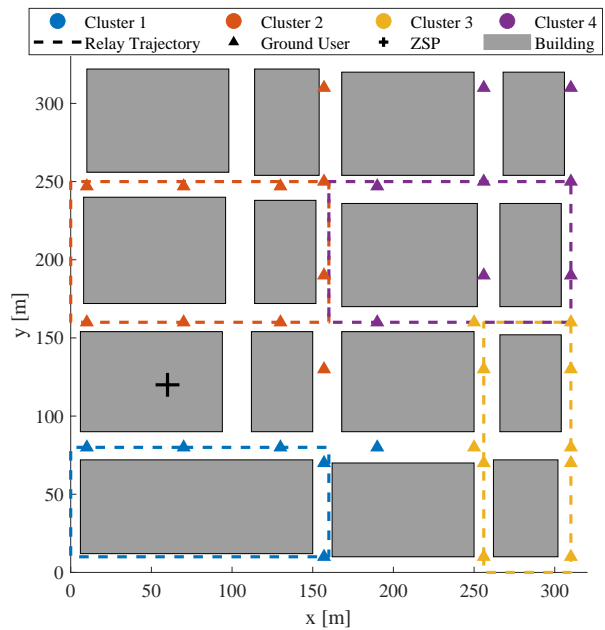}
    \caption{Scenario \#3 simulation environment.}
    \label{fig:scenario-3-map}
\end{figure}
The xy coordinates (i) are extracted from OpenStreetMap with the aid of OpenCV \cite{6240859}, (ii) rescaled according to their real profile, and (iii) transposed into the spatial reference system of the simulator. Finally, the buildings' heights are generated using a random variable uniformly distributed in $[24, 30]$, which corresponds to the characteristic height (in meters) of the buildings in that area.
As shown in Figure \ref{fig:scenario-3-map}, four \acp{GU} clusters of different size are present on the ground. Each of them is served by a drone, which relays the traffic by means of the NAT application discussed in Section \ref{sec:relayapp}.
The entire simulation lasts $\si{180}{s}$ and employs the \texttt{ns3::HybridBuildingsPropagationLossModel} to take into account the fading caused by the presence of buildings. It includes a combination of Okumura-Hata model and COST231 for long-range communications, ITU-R P.1411 for short-range communications, and ITU-R P.1238 for indoor ones. This allows to support a wide range of frequencies spanning from 200 MHz up to 2600 MHz. Moreover, each building is characterized by a window per room and is assumed to be built with concrete walls. The Wi-Fi stack has been configured based on the 802.11ax standard operating at 2.4 GHz and is controlled by the \texttt{ns3::IdealWifiManager}, which allows to keep track of the \ac{SINR}. Thanks to this mechanism, it is possible to always choose the best transmission mode to be used, i.e., a combination of modulation, coding scheme, and data rate.
\begin{figure}[!ht]
    \centering
    \includegraphics[width=\columnwidth]{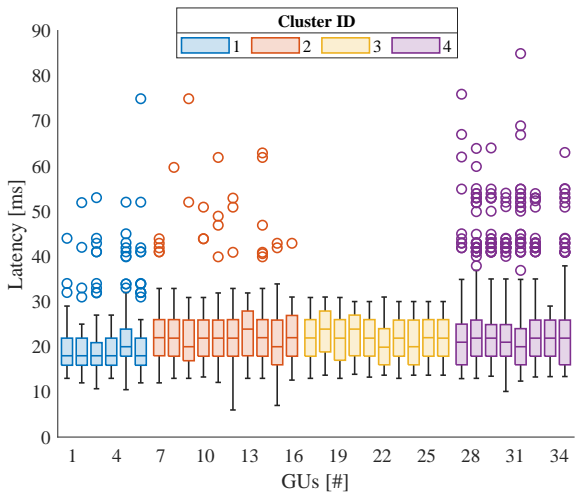}
    \caption{\acp{GU} application latency of link combined by Wi-Fi, relay drone, and \ac{LTE}.}
    \label{fig:scenario-3-latency-lte+wifi}
\end{figure}

\begin{figure}[!ht]
    \centering
    \includegraphics[width=\columnwidth]{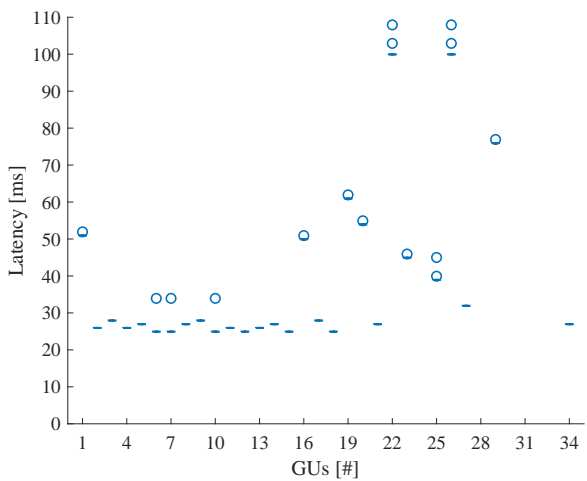}
    \caption{\acp{GU} application latency over \ac{LTE}-only link.}
    \label{fig:scenario-3-latency-lte}
\end{figure}
As for the network level, each cluster is connected to its relay according to the $10.[1-4].0.0/24$ network address range, while \ac{LTE} uses $7.0.0.0/8$.
Drones' trajectories are designed to the layout of the streets in order to minimize the shadowing effects and maximize the \ac{LoS} with the \acp{GU}. Furthermore, the path also maximizes energy efficiency as the translation in the xy plane is less costly when compared to changes of altitude. At each angle of the trajectory, the drones pause for $\si{1}{s}$ in order to simulate an accurate 90 degrees yaw.

Accordingly, each relay drone flies at a constant altitude of $\si{50}{m}$ at $\si{5}{m/s}$. Drones are equipped with the \texttt{ns3::NatApplication}, which implements a simple Port-based NAT strategy for UDP communications. Each \acp{GU} has a constant position and is equipped with a simple \texttt{ns3::UdpEchoClientApplication}, which periodically sends a packet of 1024 bytes to the remote address \texttt{200.0.0.1:1337} with a frequency of $\si{10}{Hz}$. Each packet is equipped with an application header that reports an incremental sequence number and the time of creation. Finally, the remote has a \texttt{ns3::DroneServerApplication}, which records via log messages the received packets.
\begin{figure}[!ht]
    \centering
    \begin{subfigure}[h]{.5\columnwidth}
        \centering
        \includegraphics[width=\columnwidth]{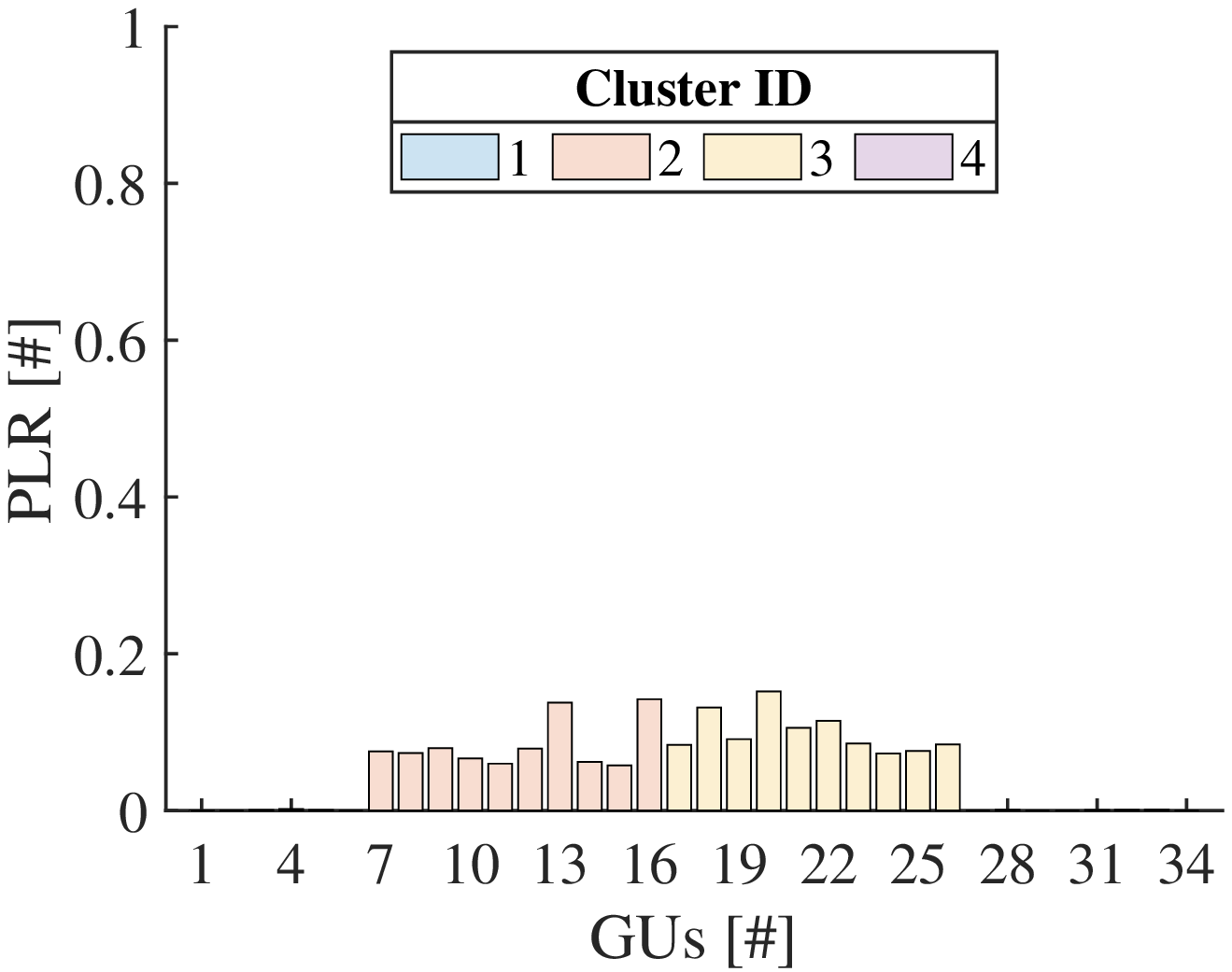}
        \caption{Wi-Fi \& LTE}
        \label{fig:scenario-3-plr-lte+wifi}
    \end{subfigure}%
    \begin{subfigure}[h]{.5\columnwidth}\hfill
        \centering
        \includegraphics[width=\columnwidth]{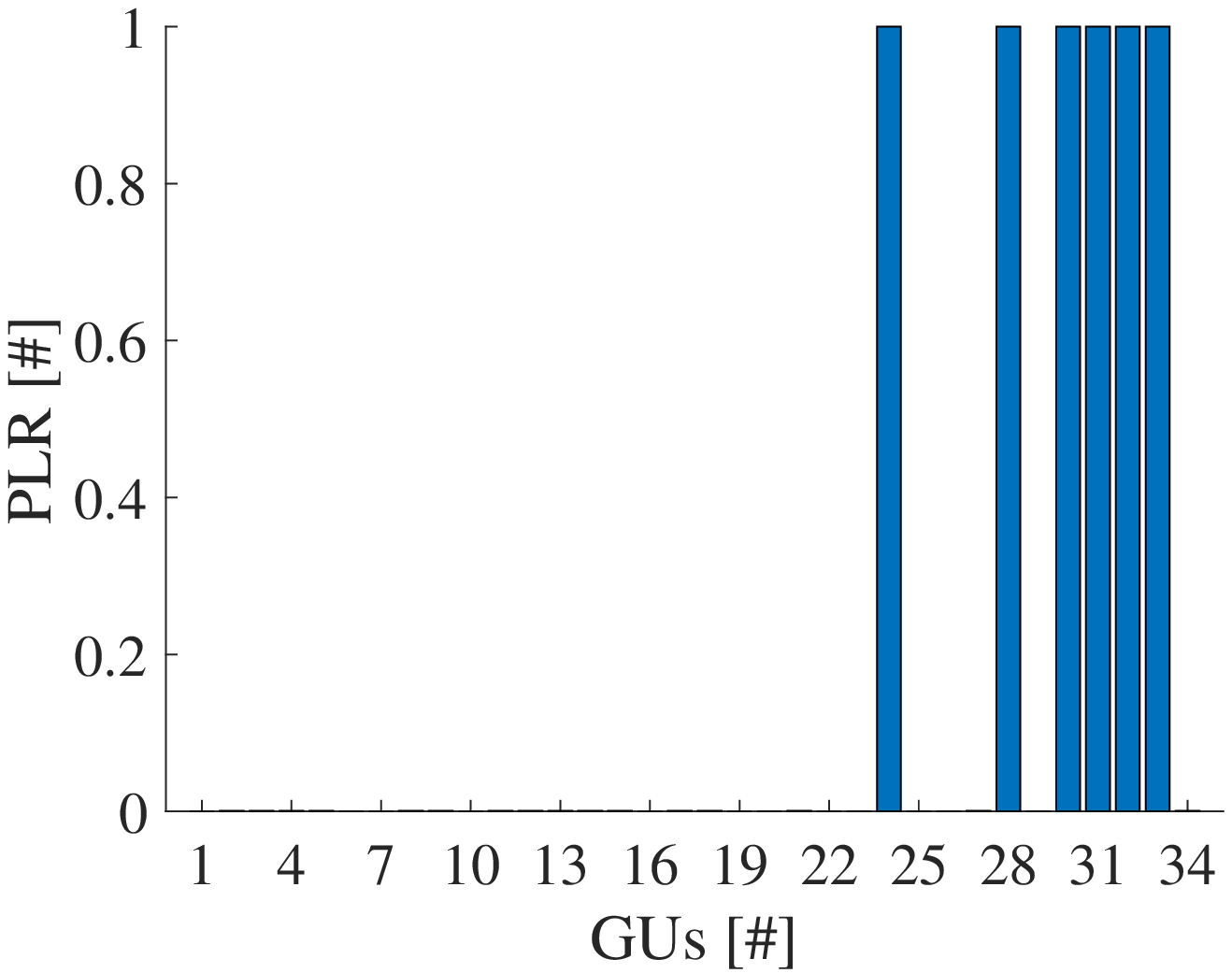}
        \caption{\ac{LTE}}
        \label{fig:scenario-3-plr-lte}
    \end{subfigure}%
    \caption{\acp{GU} application \ac{PLR} for Scenario \#3.}
    \label{fig:scenario-3-plr}
\end{figure}

\begin{table}[htbp]
    \centering
    \begin{tabular}{lrrr}
        \toprule
        Scenario \# & Events [\#] & Real Time [s] & Sim. Time [s] \\
        \midrule
        1 & 57,437 & 9 & 50 \\
        2 & 18,226,323 & 761 & 250 \\
        3 LTE & 37,178,812 & 4,620 & 180 \\
        3 Wi-Fi \& LTE & 28,903,306 & 2,858 & 180 \\
        \bottomrule
    \end{tabular}
    \caption{Comparison of the total number of events, the real time taken to execute, and the simulated time of each scenario.}
    \label{tab:perf-comp}
\end{table}
The only \ac{ZSP}, located at $[60, 120, 40]^T$, provides \ac{LTE} access to the drones, thus allowing the communication with the remote host. Figures \ref{fig:scenario-3-latency-lte+wifi} and \ref{fig:scenario-3-latency-lte} clearly show the advantage brought by the relay activity by the drones. In the relay case (Figure \ref{fig:scenario-3-latency-lte+wifi}), all the \acp{GU} experience an average latency of $\sim\si{25}{ms}$, a result that is achieved also thanks to the proposed trajectory design.

On the contrary, in absence of relay drones (see Figure \ref{fig:scenario-3-latency-lte}), while the \acp{GU} that are closer to the \ac{ZSP} are affected by a latency similar to the previous case, the farther ones register a significant delay, which inevitably compromises the reliability of the link and, hence, the \ac{QoS}. Nevertheless, this comes with a trade-off as highlighted in Figure \ref{fig:scenario-3-plr}, which shows the \ac{PLR} in both cases.
In the former, all nodes are able to transmit data to the remote, but with a loss ratio of $\sim$10\% for the cluster \#2 and \#3.
It is worth noting that this result can be further improved by properly optimizing the trajectory design to target the desired trade-off.
In the latter, instead, six nodes have 100\% \ac{PLR}, which means that there is no exchange of data.

\subsection{Performance Evaluations}
\begin{figure}[!t]
    \centering
    \begin{subfigure}[h]{.5\columnwidth}
        \centering
        \includegraphics[width=\columnwidth]{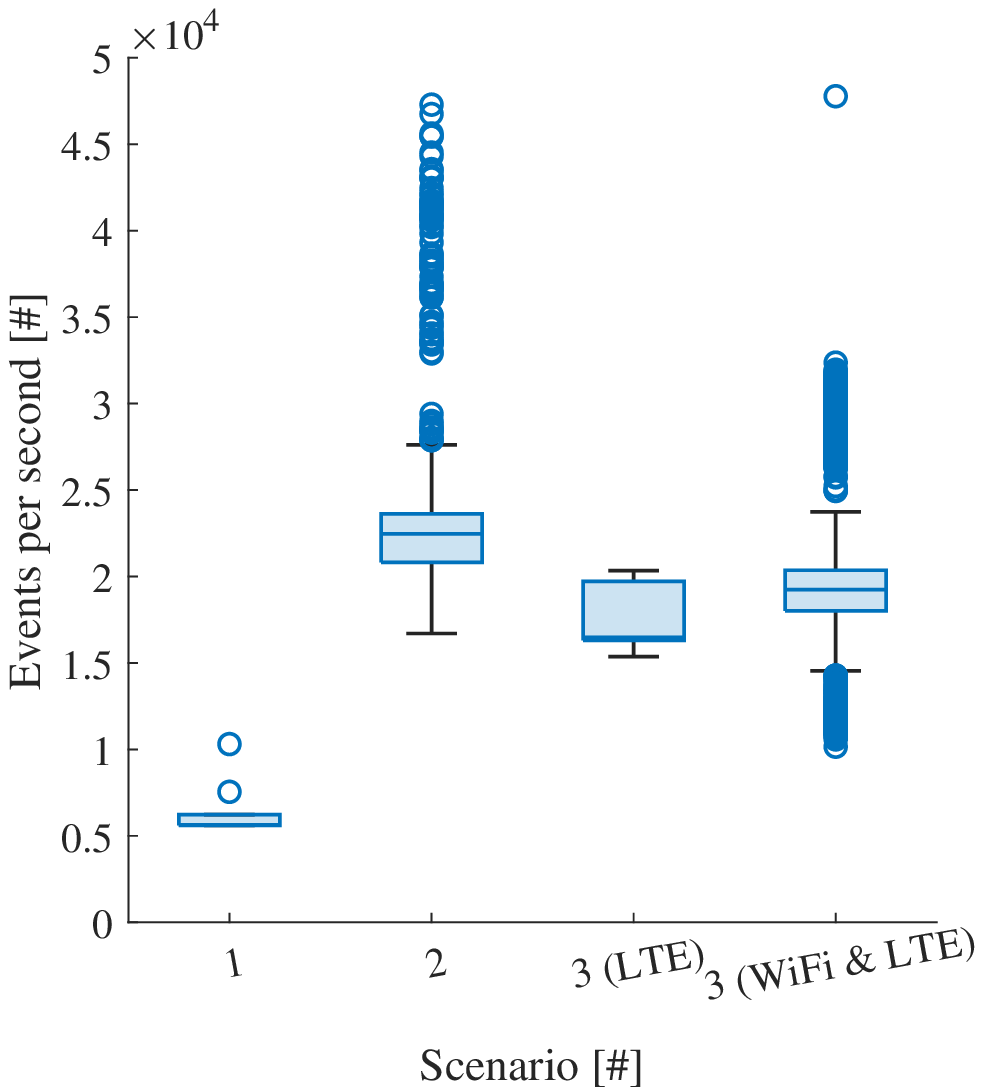}
        \caption{Number of events.}
        \label{fig:perf-events}
    \end{subfigure}%
    \begin{subfigure}[h]{.5\columnwidth}\hfill
        \centering
        \includegraphics[width=\columnwidth]{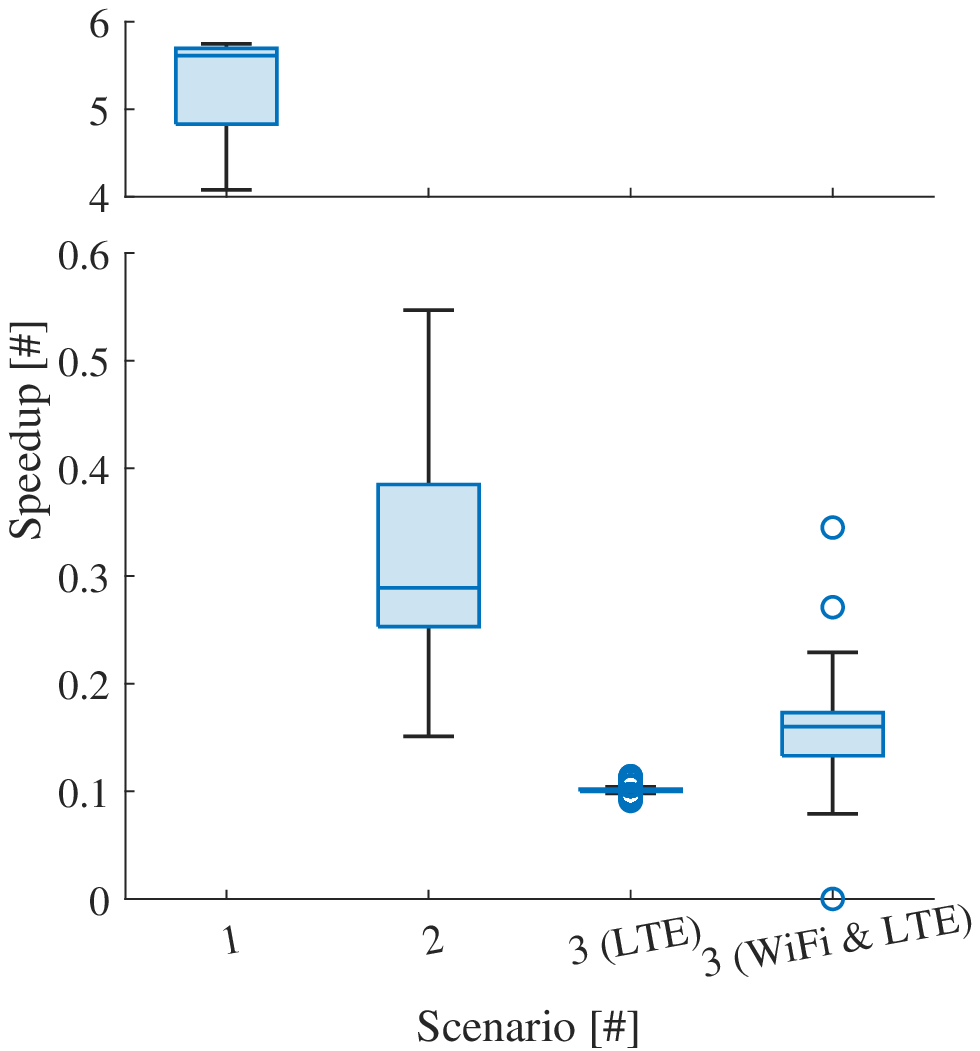}
        \caption{Simulation speedup time.}
        \label{fig:perf-speedup}
    \end{subfigure}%
    \caption{Performance evaluation of the different simulated scenarios.}
    \label{fig:perf}
\end{figure}

To evaluate the performance of the simulator, and hence its scalability, the performance metrics of the simulated scenarios are analyzed and compared hereby. The runtime environment is characterized by the following hardware and software specifications: (i) Intel (R) Xeon (R) Bronze 3106 at 1.70 GHz with 16 cores and no hyper-threading, (ii) RAM 92 GB DDR4 at 2666 MHz, (iii) 7200 RPM hard drives and (iv) OS Fedora 35 on LXD container \cite{senthil2017practical}. To fairly compare the simulations, two metrics are selected. The former takes into account the number of events processed per second for each simulation, thus providing an insight related to the scenario complexity. The latter considers the ratio between the simulated time and the real time, thus further addressing the complexity of the designed missions. Moreover, Table \ref{tab:perf-comp} summarizes the total number of events, the time taken to simulate (Real Time), and the simulated time of each scenario. It is worth noting that all scenarios are constructed differently and hence are difficult to compare. However, some clear indications can be derived from the following analysis. Indeed, Figure \ref{fig:perf} shows that in the Scenario \#1 the employment of the Wi-Fi technology slows the number of events processed per second, which means that the complexity is higher. On the contrary, the adoption of LTE (either mixed with Wi-Fi) reduces the overall computational complexity.
However, in the first case (Scenario \#1) the speedup is greater with respect to the second case (remaining scenarios): this is due to the fact that the number of generated events is way lower. This is particularly evident in the Scenario \#3, where the simulation time and the number of \acp{GU} are the same, as shown in Table \ref{tab:perf-comp}. Overall, even if the number of actors increases when drones relay are employed (LTE \& Wi-Fi), the lower number of events generated guarantees better performances.

\section{Conclusion}\label{sec:conclusions}
The \ac{IoD} paradigm enables trailblazing applications, as the flexibility proposed by drones may significantly boost the effectiveness of existing activities, e.g. first response, monitoring, delivery, and surveillance. Moreover, drones are already becoming pervasive in several industrial sectors, such as smart agriculture, proactive maintenance, civil engineering, and many more.

As a matter of fact, the large scale adoption should be evaluated after a prototyping phase that can be time consuming and may require unfeasible costs. To tackle this problem, simulators are an essential tool to facilitate the testing phase and state the readiness for real world exploitation. At the same time, simulators can be a learning tool for young professionals, engineering students and researchers to improve their knowledge and explore scenarios never considered before.

In these regards, \ac{IoD-Sim} is a thorough and user-welcoming tool that can be used to evaluate the many facets of \ac{IoD} scenarios, including trajectory design, networking functionalities, mechanical characteristics, and data analytics.
Nevertheless, \ac{IoD-Sim} has been created as a modular tool that can be updated and upgraded as needed.
A Visual Programming Editor for \ac{IoD-Sim} has also been developed, relying on compilers' theory and tools to dynamically update its contents based on the main simulator platform, ensuring that such project can be maintained with ease in the long term.
Moreover, a predictable build environment is used to ease the installation, due to its dependencies that require careful setup and knowledge about the underlying simulator, libraries, and compilers.

Even though \ac{IoD-Sim} is a reliable solution, in the future more efforts will be focused on the improvement of the entire project, especially along the following research and development lines:
\begin{itemize}
    \item Extend the support to design scenarios using technologies such as MAVlink, satellite communications, and 5G-New Radio.
    \item Speedup in Splash compilation with the use of parallel multiprocessing and optimized algorithms.
    \item Develop interactive visual blocks to preview or design more accurate simulations in lower time.
    \item Improve the overall User Experience of the visual editor.
\end{itemize}
Finally, the birth of a thriving and empowering community in open source collaboration platforms will be crucial in assessing the future development efforts of this work.